\documentclass[lettersize,journal]{IEEEtran}
\usepackage{amsmath,amsfonts}
\usepackage{algorithmic}
\usepackage{algorithm}
\usepackage{array}
\usepackage[caption=false,font=normalsize,labelfont=sf,textfont=sf]{subfig}
\usepackage{textcomp}
\usepackage{stfloats}
\usepackage{url}
\usepackage{verbatim}
\usepackage{graphicx}
\usepackage{cite}
\usepackage[capitalize]{cleveref}
\usepackage{multirow}
\usepackage{booktabs}

\begin{document}

\title{Unified Cross-modal Translation of Score Images, Symbolic Music, and Performance Audio}

\author{Jongmin Jung\IEEEauthorrefmark{1}, Dongmin Kim\IEEEauthorrefmark{1}, Sihun Lee, Seola Cho, Hyungjoon So, Irmak Bukey, \\ Chris Donahue, and Dasaem Jeong\IEEEauthorrefmark{2}

\thanks{Jongmin Jung, Dongmin Kim, Sihun Lee are with Department of Artificial Intelligence, Sogang University, Seoul, South Korea. Seola Cho is with Sogang Future Lab, Sogang University, Seoul, South Korea. Hyungjoon So is with Department of Physics Education, Seoul National University, Seoul, South Korea. Irmak Bukey and Chris Donahue are with Computer Science Department, Carnegie Mellon University, Pittsburgh, PA, United States. Dasaem Jeong is with Department of Art \& Technology, Sogang University, Seoul, South Korea.}

\IEEEauthorrefmark{1}These two authors contributed equally.

\thanks{\IEEEauthorrefmark{2}Corresponding author: dasaemj@sogang.ac.kr}

\thanks{This work was supported by the Ministry of Education of the Republic of Korea and the National Research Foundation of Korea (NRF-2024S1A5C3A03046168).}
}

\markboth{Journal of \LaTeX\ Class Files,~Vol.~14, No.~8, August~2021}%
{Shell \MakeLowercase{\textit{et al.}}: A Sample Article Using IEEEtran.cls for IEEE Journals}


\maketitle

\begin{abstract}
Music exists in various modalities, such as score images, symbolic scores, MIDI, and audio. Translations between each modality are established as core tasks of music information retrieval, such as automatic music transcription (audio-to-MIDI) and optical music recognition (score image to symbolic score). However, most past work on multimodal translation trains specialized models on individual translation tasks. In this paper, we propose a unified approach, where we train a general-purpose model on many translation tasks simultaneously.
Two key factors make this unified approach viable: a new large-scale dataset and the tokenization of each modality.
Firstly, we propose a new dataset that consists of more than 1,300 hours of paired audio-score image data collected from YouTube videos, which is an order of magnitude larger than any existing music modal translation datasets.
Secondly, our unified tokenization framework discretizes score images, audio, MIDI, and MusicXML into a sequence of tokens, enabling a single encoder–decoder Transformer to tackle multiple cross-modal translation as one coherent sequence-to-sequence task. 
Experimental results confirm that our unified multitask model improves upon single-task baselines in several key areas, notably reducing the symbol error rate for optical music recognition from 24.58\% to a state-of-the-art 13.67\%, while similarly substantial improvements are observed across the other translation tasks.
Notably, our approach achieves the first successful score-image-conditioned audio generation, marking a significant breakthrough in cross-modal music generation. 
\end{abstract}



\section{Introduction}
\IEEEPARstart{A}{} piece of music can be represented in various forms, each with distinct characteristics. For example, a Beethoven piano sonata would typically be distributed as printed scores, where the composer notates their musical ideas with symbols; these scores may exist as scanned or synthesized images, or in machine-readable notation formats such as MusicXML. Specific performances of the piece can be recorded as audio files but can also be represented in time-aligned symbolic formats such as MIDI, where each note’s onset and offset timing is explicitly provided.

\begin{figure}[t]
    \begin{center}
    \centerline{\includegraphics[width=\columnwidth]{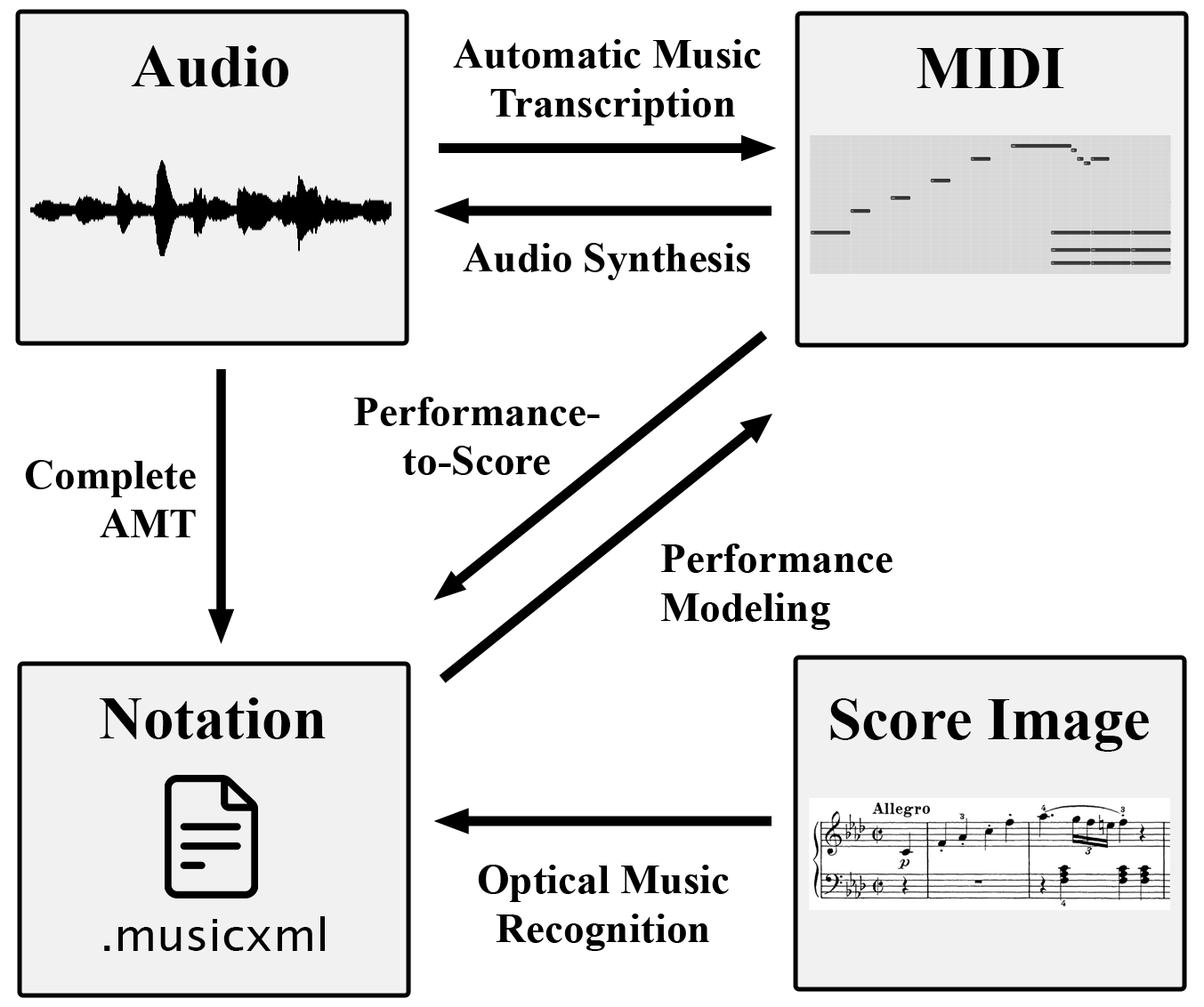}}
    \caption{Conventional cross-modal conversion tasks in music information retrieval research. 
    }
    \label{figure-mir-tasks}
    \end{center}
\end{figure}

As each modality serves a different purpose, converting music from one representation to another has garnered much research interest in the field of music information retrieval (MIR) and established numerous mainstay tasks: as illustrated in \cref{figure-mir-tasks}, automatic music transcription (AMT)~\cite{amtsurvey, amtsurvey2}, MIDI-to-audio synthesis~\cite{hawthorne2018maestro}, optical music recognition (OMR)~\cite{omrsurvey2, omrsurvey}, complete music transcription~\cite{nakamura2018towards}, performance modeling~\cite{jeong2019graph}, and performance-to-score conversion~\cite{Liu2022PerformanceMC, beyer2024end}.

However, previous research has largely treated these tasks as separate problems, relying on specialized datasets and methodologies designed for each task. Some works have explored multi-modal pipelines---first converting music notation into performance MIDI and subsequently synthesizing audio \cite{dong2022deep, tang2025towards}---but typically through multi-stage frameworks using separate models.

We can instead consider a task that inherently encompasses multiple modal translations at once. The most extreme case of this is score image to performance audio conversion—--the process of directly generating music audio from a score image, bypassing symbolic notation altogether.
To the best of our knowledge, this task has never been attempted before, likely because it demands mastery of both OMR and controllable music generation, two open areas of research.
Yet, it mirrors how human musicians interpret music: when sight-reading, performers directly translate score images into expressive performances, without explicitly constructing an intermediate symbolic representation.

In this paper, we propose a unified multimodal music translation framework, based on a new large-scale dataset and a unified modeling approach. Following the previous research in unifying image-language-speech modal translation~\cite{kim2024tmt}, we formulate music modal translation as a seq2seq task using vector-quantized discrete tokens of image and audio. This unified framework enables a synergistic interaction between different modal translation tasks, where information learned from one task can benefit others, even when the input and output modalities do not directly overlap. Specifically, our result shows that OMR performance can be improved when we combine MIDI-to-audio synthesis task. 

\begin{figure*}[ht]
    \begin{center}
    \centerline{\includegraphics[width=\textwidth]{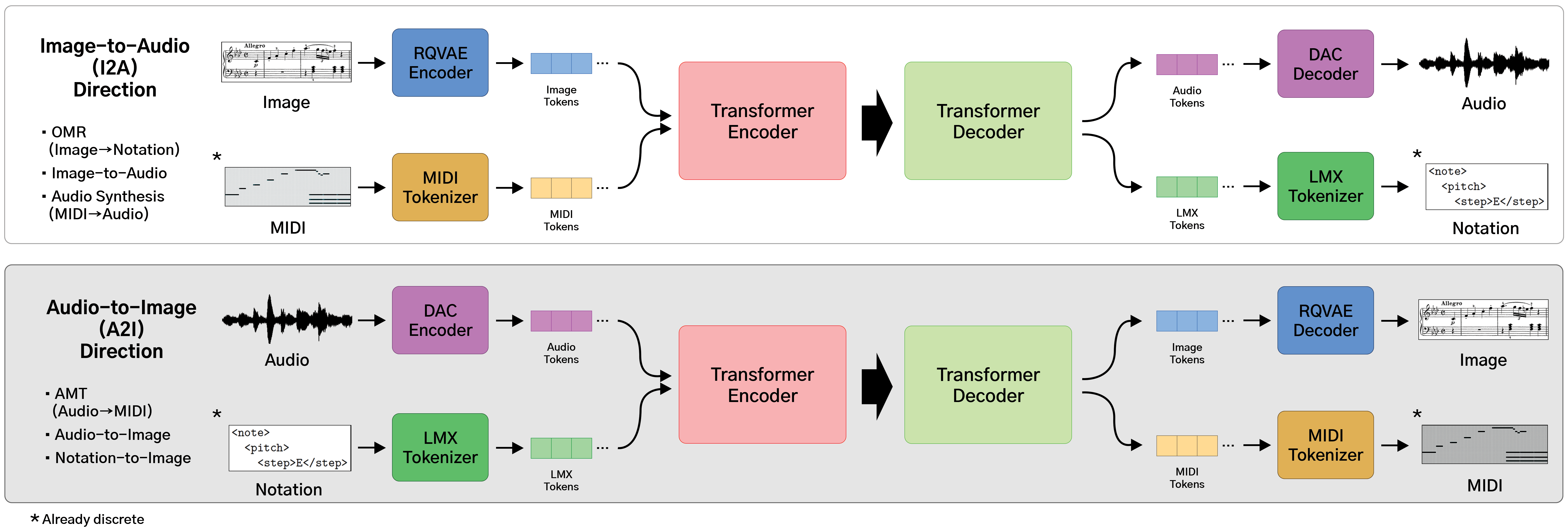}}
    \caption{
    Overview of our proposed unified multimodal translation framework.
    We employ a single Transformer encoder-decoder model for each direction—
    one for \textit{Image-to-Audio direction} (I2A) tasks and another for \textit{Audio-to-Image direction} (A2I) tasks.
    Each model jointly handles multiple translation tasks.
    All modalities are discretised into token sequences, enabling end-to-end, multitask training entirely at the token level.
    Note that we train separate models for I2A and A2I directions; the two directions do not share weights.
    }
    \label{figure-model}
    \end{center}
\end{figure*}

While the image-audio pairs do not include any symbolic information, such as MusicXML or MIDI, their musical semantics are closely connected.  To generate audio from a given score image, a model must implicitly handle all modal translations: recognizing notes and instructions from score images, predicting the timing and dynamics of each note, and synthesizing the notes into audio. 
Conversely, for audio-to-image generation, the model must identify the notes from audio, organize them into formalized notation, and engrave the notation in sheet music form.
Based on our experiments, we show that learning outer modality translations (image $\leftrightarrow$ audio) can also enhance performance for inner modality tasks such as OMR and AMT.

Furthermore, we find that our image-to-audio and audio-to-image translation models function properly only when trained alongside other modal tasks that align with their translation direction (as illustrated in \cref{figure-model}). This highlights that the unification of modal translation tasks is not just an optional enhancement but a necessary component in making score-to-audio or audio-to-score generation feasible.

Our main contributions are as follows:
\begin{itemize}
    \item We propose the first unified multimodal music translation method that covers the four most representative modalities of Western music---score images, symbolic scores, performance MIDI, and audio---in a single model.
    \item We are the first to successfully train a model that can generate music audio given a score image, closing the gap with the capabilities of human musicians.
    \item We introduce the YouTube Score Video (YTSV) dataset that consists of more than 1,300 hours of score images paired with corresponding performance audio recordings.
    \item We demonstrate that the combined multimodal training can enhance the performance of subtasks. 
\end{itemize}


\section{Problem Formulation and Related Work}
\label{sec:problem-formulation}

Before detailing our methodology, it is essential to clarify the distinct modalities of music representation that our framework aims to bridge. In this work, we consider the following four modalities of
music representation:

\begin{itemize}
    \item \textbf{Score Image}: Sheet music in the form of raw images. Typically scanned from real-world prints or rendered with digital notation software. We cover both synthetic and scanned images while discarding hand-written scores.

    \item \textbf{Symbolic Notation}: Digitized semantic information of sheet music, such as MusicXML (MXL). Along with the pitch and duration of notes, it also includes detailed elements such as slurs, voicing, articulation, and tempo and dynamic markings that are essential to render human-readable music scores. We employ a modified version of the industry-standard format MusicXML, which we describe further in \cref{lmx}. 

    \item \textbf{MIDI}: Event-based symbolic representation of the performed notes with concrete note onset and offset timings. While notational details such as slurs or articulation are omitted, MIDI is better suited to express human performance with expressive tempo and dynamic changes.

    \item \textbf{Audio}: Recorded from real-world performance or synthesized with virtual instruments. 
\end{itemize}

\begin{figure}[t]
    \begin{center}
    \includegraphics[width=\columnwidth]{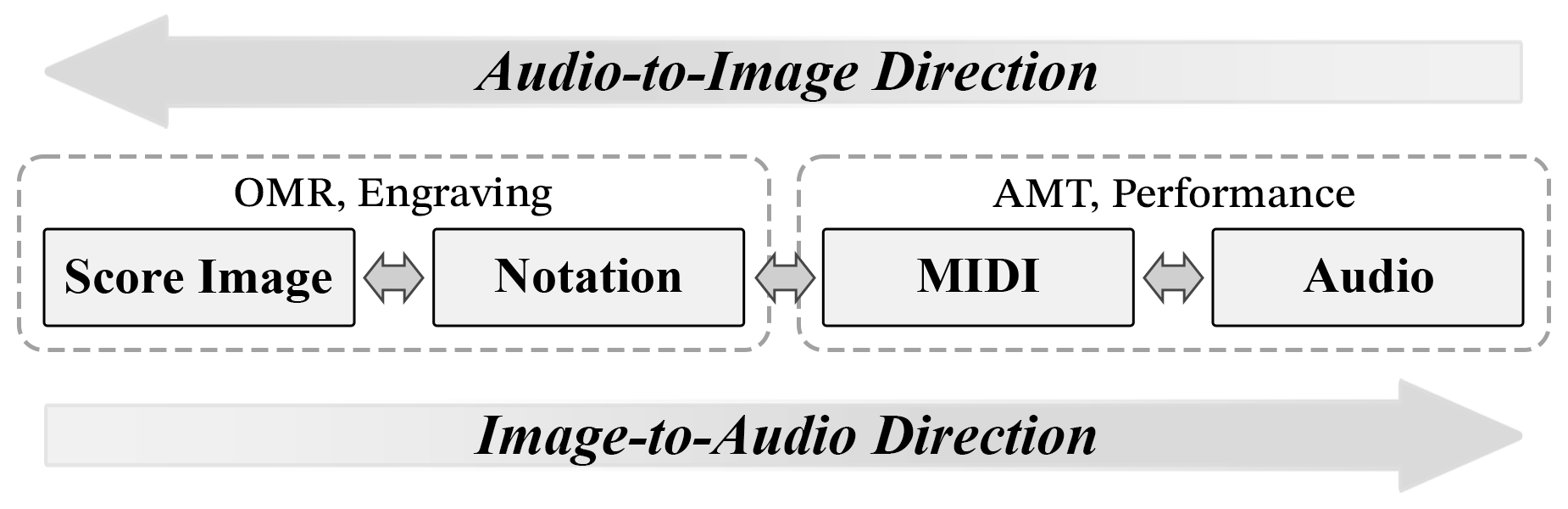}
    \caption{The four modalities of music representation used in this paper.}
    \label{figure-modals}
    \end{center}
\end{figure}

As shown in \cref{figure-modals}, we observe that these four representations exist on a ``modal spectrum''; each one naturally leads to the next in the process of music-making and in MIR tasks. 
As the modalities have a strong causal relationship with one another and each highlight different aspects of music, the information learned for one subtask can have a synergistic effect with others.
We consider translation tasks starting from the score image-side as being in the \emph{Image-to-Audio direction} (\textbf{I2A}), and translation tasks starting from the audio-side as being in the \emph{Audio-to-Image direction} (\textbf{A2I}).
Among many possible modal translations, we mainly focus on automatic music transcription and optical music recognition, the key representative tasks in each direction that have been actively researched. To along with these, we also introduce direct image-to-audio and audio-to-image translation as as novel challenging tasks that expand the boundaries of cross-modal music translation beyond adjacent conversions, addressing the end-to-end capabilities of the translation models.

\subsection{Automatic Music Transcription}
Automatic music transcription (AMT) is the task of inferring musical notes or notations from audio, and is considered one of the fundamental problems in MIR. The output of AMT can range anywhere from simple piano-roll representations to structured sheet music data~\cite{BenetosAMToverview}.
Machine learning-based AMT has seen much success in recent years, particularly with piano music \cite{hawthorne2017onsets, kong2021high}, thanks to the vast amount of available training data with high precision labels collected with computer-controlled pianos~\cite{hawthorne2018maestro}. While piano-roll-like frame-wise prediction has been widely used for AMT models, Hawthorne \emph{et al.} \cite{hawthorne2021sequence} introduced the first AMT model utilizing token-based prediction using a transformer encoder-decoder structure. This approach has been further adapted to multi-instrument AMT~\cite{gardner2021mt3, chang2024yourmt3+}.

\subsection{Optical Music Recognition}
The aim of optical music recognition (OMR) is to transcribe music notation from score images. While previous methods typically involve multi-stage processes, recent state-of-the-art approaches adopt an end-to-end strategy that takes images as input and predicts token sequences for the corresponding symbolic notation~\cite{rios2024sheet, mayerZeus}. Due to the difficulty in handling complex score layouts, it was only recently that OMR research on polyphonic piano form score images had seen successful results~\cite{rios2023end}. 

A significant bottleneck in OMR research is the limited availability of labeled training data in various styles; while many resources exist for machine-readable symbolic music notation, corresponding real-world scans are relatively scarce. Therefore, much of OMR research relies on synthetic datasets for model training, while reserving scanned images for evaluation~\cite{mayerZeus, rios2024sheet2}.


\section{Methods}

\subsection{Multimodal and Multitask Approaches}
Recent works have explored unified models that tackle multiple tasks and modalities simultaneously. UniAudio\cite{pmlr-v235-yang24x} introduced a single model trained on 11 diverse audio generation tasks, spanning inputs from audio, text, MIDI, and phonetic sequences, by leveraging a discrete audio token representation. Fugatto\cite{valle2025fugatto} likewise demonstrated that a unified model conditioned on both audio and text can achieve remarkably fluent results across various audio generation tasks. Another notable attempt was proposed by Kim \emph{et al.}~\cite{kim2024tmt}, which trains a single sequence-to-sequence Transformer on discrete token sequences to translate between images, text, and speech. While these prior works showcase the benefits of multitask training with unified representations, they mostly target well-studied tasks on established datasets and focus on a limited range of modalities. 

In contrast, our work extends the unified multimodal paradigm to the domain of music translation across score and performance modalities, introducing new translation tasks and addressing data challenges. Specifically, our proposed model handles a broader set of output modalities beyond audio tokens: in addition to the traditional tasks of automatic music transcription (AMT) and optical music recognition (OMR), we enable the novel cross-modal generation of performance audio from sheet music images, and vice versa (generating sheet images from audio). To mitigate the data scarcity in tasks like AMT and OMR, we contribute a new large-scale dataset (over 1,300 hours of aligned score-image/audio pairs). Even though these image–audio pairs lack explicit note-level alignments, we demonstrate that incorporating them into a unified multitask framework allows the model to implicitly learn shared musical structure, leading to improved performance on transcription tasks. 

\subsection{Tokenization}\label{sec:tokenization} 

To enable our unified sequence-to-sequence modeling across diverse music modalities, we convert every input and output into a sequence of discrete tokens. Continuous data modalities such as score images and audio are discretized using learned neural compression models, which quantize the raw image pixels or audio waveform into compact token sequences. In contrast, symbolic modalities like MusicXML scores and MIDI performances are tokenized by structurally flattening their hierarchical representations into linear sequences of musical tokens. This unified tokenization scheme provides a common sequential format for all modalities and is essential for training a single Transformer that can handle the full range of cross-modal music translation tasks.

\subsubsection{Image and Audio Tokens}

We train two discrete representation models: Residual Quantized Variational Autoencoder (RQVAE)~\cite{rqvae} and Descript Audio Codec (DAC)~\cite{dac} for images and audio, respectively. Prior to tokenization, we threshold the images by identifying the median pixel value of each image and setting all pixels above \(median - 20\) to white (255), reducing noise and enhancing contrast; and we resample all audio to 44.1kHz and convert them to mono.

We train RQVAE with 16x compression, four unshared codebooks each containing 1,024 codes, and a model dimension of 256. Training images were randomly cropped with multiple resolutions. We do not include attention blocks in the RQVAE to ensure that the model only uses local information. We implemented a vertical flattening method for 2D image token sequences to match how sheet music is naturally read. Therefore, the model processes music vertically first (top-to-bottom), then moves horizontally right to the next column, mirroring standard sheet music reading patterns (left-to-right).

We train DAC with a token hop size of 512 samples on our 44.1kHz source material, resulting in approximately 86.13 RVQ token sets per second. The model employs four unshared codebooks with 1,024 codes each and an embedding size of 1,024. We retrain DAC because the publicly available model uses nine codebooks, which is unnecessarily large for our domain. By limiting to four codebooks and training specifically on classical music sounds rather than diverse audio types, we aim to achieve richer representations within a narrower domain while maintaining high reconstruction quality.

During tokenization, we observe that even slight shifts in pixel positions or audio samples can result in vastly different token assignments, a limitation inherent to discrete token representations. To address this and enhance model robustness, we implement comprehensive augmentation strategies: for images, we generate 32 variants through combinations of 8 horizontal and 4 vertical pixel shifts, and for audio, we create 9 variations through temporal shifts.

\subsubsection{Linearized MusicXML}
\label{lmx}

MusicXML is a machine-readable musical notation format for storing semantic information of sheet music. While the format's versatility has led to a wide adoption in editing software and music processing libraries, its hierarchical structure and verbosity pose challenges for processing with language models. To mitigate this, Mayer \emph{et al.} \cite{mayerZeus} introduced \textit{Linearized MusicXML (LMX)}, a modification of the format that is much more concise and a better fit for token-based generation. We use the original implementation of LMX to represent the notation data in all of our experiments.

\subsubsection{MIDI-Like Tokens}

MIDI represents musical performance as a sequence of digital events, capturing note information and timing. While this format effectively encodes performance data, its event-based structure requires adaptation for language model processing. Drawing from the MT3 framework \cite{gardner2021mt3}, we adopt a MIDI-like tokenization scheme that transforms MIDI data into discrete tokens that include instrument identifiers, MIDI pitches, note on/off events, and quantized time markers at 10ms intervals. For this, we use the specific implementation introduced in YourMT3+~\cite{chang2024yourmt3+}.

\subsection{Model Architecture and Unified Token Space}
All data inputs and outputs are first converted into sequences of discrete tokens, using modality-specific tokenizers. This yields a unified token space in which each modality $\mathcal{X}\in{\{\mathcal{I, A, N, M}\}}$ (Image, Audio, Notation, MIDI) has its own vocabulary of tokens $\mathcal{V_X}$. We augment each modality’s vocabulary with special start-of-sequence and end-of-sequence tokens (\texttt{[SOS]}$_\mathcal{X}$, \texttt{[EOS]}$_\mathcal{X}$), and use a separator token \texttt{[SEP]} to mark boundaries (e.g., between musical systems in an image) in the image domain. The overall vocabulary $\mathcal{V}$ is the union of all modality-specific token sets and these special tokens. 

A single sequence-to-sequence encoder-decoder Transformer \cite{vaswani2017attention} model is then used to perform the translation between token sequences. The encoder takes as input the source token sequence $z^{(\mathcal{X})}_{1:L_X}$ (of length $L_X$) and maps it to a sequence of hidden representations; the decoder autoregressively generates the target sequence $z^{(\mathcal{Y})}_{1:L_Y}$ token by token. To inform the model of the desired target modality, we add a target-modality embedding to every input token representation. Formally, the input embedding for token $x_i$ from source modality $\mathcal{X}$ when the target is modality $\mathcal{Y}$ is $e_i = \text{TokEmb}(z^{(\mathcal{X})}_i) + \text{PosEmb}_\mathcal{X}(i) + \text{TgtEmb}_\mathcal{Y}$, where $\text{PosEmb}_\mathcal{X}(i)$ is a modality-specific learnable positional embedding and $\text{TgtEmb}_\mathcal{Y}$ is a trainable vector representing the target modality $\mathcal{Y}$. This scheme allows the same model to handle any source–target modality pair with a single set of parameters. More details are given in Appendix~\ref{sec:math_formulation}.

For image and audio outputs, which are represented by multiple parallel codebooks of tokens (from RQVAE and DAC), we employ a specialized decoding approach so that the Transformer does not need to output all codebook tokens as one long flattened sequence. In particular, we follow the codebook-wise decoding strategy of Yang \emph{et al.}~\cite{pmlr-v235-yang24x}, by augmenting the Transformer decode with an additional one-layer sub-decoder module that produces codebook tokens in parallel for each step. This enables generating the multiple codebook streams for images or audio in a fully causal fashion without flattening them into a single sequence. 

We train the model using the standard cross-entropy loss to maximize the likelihood of the target token sequence given the source sequence. The training used teacher-forcing strategy, and at inference, the model produces outputs autoregressively. This unified model design allows a single Transformer to learn all modal translations jointly in a multi-task fashion under a consistent tokenization scheme.


\section{YouTube Score Video Dataset}

\begin{figure}[t]
    \begin{center}
    \centerline{\includegraphics[width=\columnwidth]{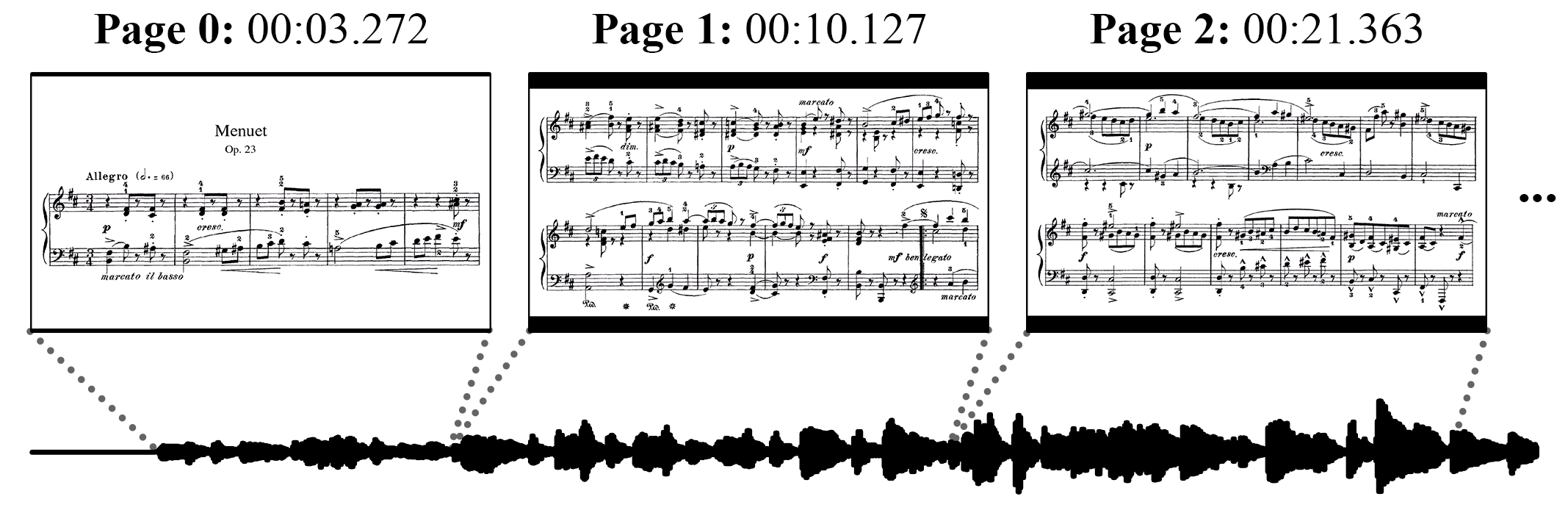}}
    \caption{An example from one of the videos collected for the YouTube Score Video dataset. Slides of sheet music are aligned to the corresponding points in audio.}
    \label{figure-video}
    \end{center}
\end{figure}

While there has been numerous datasets aligning multimodal music data using manual annotation or automatic audio-score alignment on either symbolic music score \cite{weiss2021schubert, weiss2023wagner, thickstun2016musicnet} or sheet image \cite{yang2019MIDI, feffer2022assistive}, these datasets aimed to annotate the alignment densely, such as measure-level. However, if sparse alignment checkpoints---such as system- or slide-level transitions---are deemed sufficient, a far greater amount of data becomes accessible. 
In particular, YouTube hosts an abundance of public \textit{score-following} videos for Western classical music, where uploaders manually divide sheet music into slides, each typically containing two or three systems (i.e., lines of music), and precisely synchronize slide transitions with performance audio. These videos are primarily intended to help viewers follow along with the score while listening and are widely used for music education and enjoyment. As illustrated in ~\cref{figure-video}, such videos inherently offer weak but musically meaningful alignment between score images and audio, providing a valuable and scalable resource for multimodal music translation.

\subsection{Data Collection}

\begin{table}[t]
    \caption{Data distribution of the YouTube Score Video dataset after filtering}
    \label{lsyt-stats}
    \begin{center}
    \begin{small}
    \begin{tabular}{l||r|r|r}
    \hline
    Category & Videos & Segments & Duration (hrs) \\
    \hline
    Solo Piano           & 9,052 & 232,029 & 762.34 \\
    Accompanied Solo     & 912 & 47,373 & 141.83 \\
    String Quartet       & 594 & 48,470 & 138.48 \\
    Others (Chamber)     & 1,659 &  106,048 & 298.65 \\
    \hline
    Total                & 12,217 & 433,920 & 1,341 \\
    \hline
    \end{tabular}
    \end{small}
    \end{center}
\end{table}

We collect 12,217 score-following videos featuring various categories of western classical music. Each image-audio pair in a video forms a single data sample. \cref{lsyt-stats} shows the statistics of the YouTube Score Video dataset.

Despite the lack of notation and MIDI data, the dataset contains a great amount of high-quality, in-the-wild sheet music and performance audio; to the best of our knowledge, it is the largest available collection of its kind.\footnote{The metadata for the dataset including links for each video, along with the codebase for preprocessing will be publicly released upon acceptance of the paper.} Here we describe the various measures we take to extract and process the data samples.

\subsection{Data Extraction and Processing}

\textbf{Slide Segmentation.} We devise a system for identifying the timing points of slide transitions in the collected score-following videos. The transitions manifest in two distinct ways: instantaneous cuts between pages, and animated transitions such as crossfades or wipes. To handle this, we develop a rule-based segmentation algorithm that accommodates both cases while maintaining temporal accuracy. This process extracts the individual slides of score images along with their corresponding audio slices, creating the foundational paired segments for the dataset. Details are elaborated in Appendix \ref{sec:slide-seg}.

\begin{figure}[t]
    \vskip -0.05in
    \centering
    \includegraphics[width=\columnwidth]{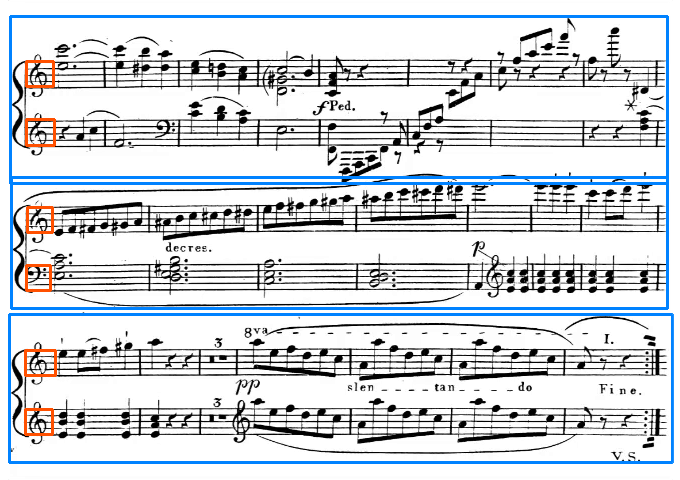}
    \vskip -0.1in
    \caption{Illustration of the music system detection pipeline using the fine-tuned YOLOv8. Music systems detected by fine-tuned YOLOv8 are notated with blue boxes, and detected staff lines are notated with red boxes. Note that the red boxes detect the staff height near clefs, not the clefs themselves. 
    }
    \label{fig:lsyolo}
\end{figure}

\textbf{System Cropping and Resizing.} The collected videos often feature letterboxes or pillarboxes, diverse aspect ratios, and inconsistent margins around contents; also, each slide contains an arbitrary number of musical systems alongside non-musical elements such as titles. To handle these irregularities, each musical system in a slide must be cropped and resized in a consistent format. 
Therefore, we label new annotations to fine-tune YOLOv8-based~\cite{yolov8} models for two different tasks: system-wise bounding box regression, and staff line detection, as depicted in \cref{fig:lsyolo}. We use the models to produce regularized crops of each system, a process which we discuss further in Appendix \ref{sec:lsyolo-system}.

\subsection{Data Filtering}
We filter the dataset in two levels: in the corpus-level, based on metadata; and in the sample-level, focusing on individual data quality.

\textbf{Corpus-level Filtering.} We employ the large language model Claude Sonnet 3.5~\cite{claude} to extract structured metadata from video titles, such as composer information, instrumentation, composition year, and category.
Based on the collected metadata, we filter out orchestral compositions, pieces including singing voices, and some larger forms of chamber music. This is to rule out pieces that include a large number of staves per system or lyric texts, which would add excessive complexity to image tokenization. Additionally, we exclude pieces newer than the year 2000, as contemporary classical music often feature unconventional or highly complex notations. We further refine our dataset by removing handwritten or annotated scores.

\textbf{Sample-Level Filtering.} We address more granular aspects of data quality with algorithmic filtering. First, we implement pixel intensity-based metrics to identify and remove poor-quality scans and visual manipulations added by the uploader; second, we discard overly long or short audio samples, which might indicate errors in slide segmentation; finally, we filter cropped images based on their size and overlap with other bounding boxes to handle potential errors in system cropping.

For model training, we only use segments that are shorter than 20 seconds in audio and smaller than 256,000 total pixels in image. After filtering, we obtained 433,920 image-audio pairs from 12,217 videos, which is 1,341 hours in total. This includes approximately 10,000 unique pieces by more than 2,000 composers.


\section{Data Collection}

The proposed multimodal, multi-task learning strategy requires a substantial amount of data in four distinct forms of music representation. Since existing datasets rarely cover all of the necessary modalities, we combine various datasets of different compositions to ensure adequate data in all four.

In addition to the YouTube Score Video dataset, we collect and process the following public datasets:

\textbf{GrandStaff Dataset} \cite{rios2023grandstaff}: a collection of synthetic pianoform \textbf{score images} and \textbf{notation}. It consists of solo piano repertoires from six composers. The original 7,661 samples were augmented with six key augmentations, yielding 53,882 samples. Additionally, visual distortions were applied to generate paired distorted versions of the scores, resulting in 107,764 score-images in total. Every score is segmented to the system-level.

\textbf{OLiMPiC Dataset} \cite{mayerZeus}: contains scanned and synthetic pianoform \textbf{score images} and \textbf{notation}. Only synthetic images are included in the training set. The notation was sourced from the Open Score Lieder Corpus~\cite{gotham2022openscore}, which is crowd-sourced transcriptions of 19th-century piano-accompanied art songs, and only the piano staves are cropped from score images. The scores are segmented at the system level, resulting in a total of 17,945 samples.

\textbf{MusicNet} \cite{thickstun2016musicnet}: dataset of classical music \textbf{audio} and corresponding time-aligned \textbf{MIDI} annotations. We use the MIDI labels from MusicNet$_{EM}$ \cite{maman22a} that provide better alignment to the audio. 

\textbf{MAESTRO}~\cite{hawthorne2018maestro}: a collection of \textbf{audio} and \textbf{MIDI} data from piano performances. MIDI data is captured with Disklaviers (computer-controlled pianos), thus providing accurate timing precision.

\textbf{SLakh} \cite{manilow2019slakh}: a dataset of synthesized multi-track \textbf{audio} for the \textbf{MIDI} data in the Lakh dataset\cite{raffel2016learning}. We only use the mixed audio and discard individual stems.

\textbf{Beethoven Piano Sonata Dataset} (BPSD) \cite{zeitler2024bpsd}: \textbf{score images}, \textbf{notation}, multi-version \textbf{audio}, and their corresponding time-aligned \textbf{MIDI} of Beethoven's piano sonatas.

As BPSD does not provide system-level alignment between score images and other modalities, we manually annotate and align our own test subset, selecting 9 pieces that belong to the MAESTRO test set. 

\begin{table}[t]
\caption{Data Distribution of Combined Datasets with Aligned Modalities.}
    \label{data-stats}
    \begin{center}
    \begin{small}
        \begin{tabular}{l||c|c|c|c||r|r}
            \hline
            \multirow{2}{*}{Subset} & \multicolumn{4}{c||}{Modalities} & \multirow{2}{*}{N} & \multirow{2}{*}{H} \\
            \cline{2-5}
                       & Img   & MXL & MIDI    & Aud   &  &  \\
            \hline
            YTSV       & $\surd$ & -        & -           & $\surd$ & 433,920  & 1,341\\
            GrandStaff & $\surd$ & $\surd$  & -           & -       & 7,661   & $^*$23 \\ 
            OLiMPiC    & $\surd$ & $\surd$  & -           & -       & 17,945   & $^*$47 \\ 
            MusicNet   & -       & -        & $\triangle$ & $\surd$ & 330   & 33 \\
            MAESTRO    & -       & -        & $\surd$     & $\surd$ & 1,276  & 199  \\
            SLakh      & -       & -        & $\surd$     & $\surd$ & 2,100  & 145 \\
            BPSD       & $\surd$ & $\surd$  & $\triangle$ & $\surd$ & 32    & 14 \\
            \hline
            \end{tabular}
    \end{small}
    \end{center}
\end{table}

\cref{data-stats} shows the composition of the combined datasets used in our experiments, where $\triangle$ denotes MIDI data made from audio-aligned scores, N denotes the number of aligned entries, and H denotes audio duration in hours. The ymbolic notation datasets notated with $^*$ do not have audio files to report the duration. Therefore, for the OLiMPiC dataset, we computed the total duration by summing the audio lengths of every piece contained in the dataset, as provided by the official Open Score Lieder Corpus account on MuseScore~\cite{GothamJonas2022}. Likewise, for the GrandStaff dataset, we obtained the total playing time by summing the MIDI durations of all pieces that constitute the dataset in KernScores~\cite{kernscores}. The collected datasets, combined with the YouTube Score Video Dataset, forms the final assemblage used for model training.


\section{Experiments}

\subsection{Modal Directions}
Although it is theoretically possible to train a single model that covers all modal translations, our preliminary experiments showed many practical difficulties. 
Instead, we train two different models in each of two directions described in cref{sec:problem-formulation}: \emph{Image-to-Audio} (\textbf{I2A}) and \emph{Audio-to-Image} (\textbf{A2I}). Each model is trained only for modal translations in its respective direction; for example, the I2A model is trained for image-to-audio, image-to-notation, and MIDI-to-audio tasks. Thus, I2A direction model has the input modality $\mathcal{X}\in\{\mathcal{I,M}\}$ and the output modality $\mathcal{Y}\in\{\mathcal{N,A}\}$, while A2I direction model has the opposite.

For the I2A model, we only focus on piano music instead of the entirety of chamber music, as modeling audio token sequences for various timbres proves challenging with the current dataset size. Also, the current OMR (image-to-LMX) dataset only covers piano repertoire, making it difficult to measure the effect of incorporating non-piano music in the other tasks; therefore, for image-to-audio translation, we use a subset of the YTSV dataset consisting of piano music (\emph{YTSV-P}, 252k segments and 815 hours of audio), and only the MAESTRO dataset for MIDI-to-audio.
For the A2I model, we instead use the entirety of the YTSV dataset.  

\subsection{Implementation}

Our Transformer model uses 12 encoder layers and 12 decoder layers, with model dimension 1024, feed-forward hidden size 4096 (with GELU activation) and 16 attention heads per layer, which are the same for both the I2A and A2I model variants. We initialize the token embedding matrix for the image and audio code tokens using the learned codebook embeddings from the RQVAE and DAC representation models, respectively, and initialize all other parameters randomly. We train each model using the AdamW optimizer \cite{loshchilov2018decoupled} with an initial learning rate of $1\times10^{-4}$. The learning rate is scheduled to decay to $1\times10^{-5}$ following a cosine decay schedule, after a linear warm-up of 2,000 steps. Training is run for 600,000 updates, with a total batch size of 24 sequence pairs, on 2× NVIDIA H100 SXM GPUs. For OMR, AMT and MIDI-to-audio tasks, the multi-task models were fine-tuned for the given task for 50k steps with a learning rate of $1\mathrm{e}{-5}$.

To handle the varying lengths of music pieces, we apply sequence-length truncation during training. For the I2A model (image-to-audio direction tasks), we randomly slice each training sample so that audio clips are at most 20 seconds long (1723 tokens), and MIDI sequences are at most 1000 tokens long. For the A2I model (audio-to-image direction), we use shorter truncations (10-second audio segments, up to 1000 MIDI tokens) to account for the more difficult audio-input tasks. 

We also employ a curriculum learning strategy to introduce the different types of task gradually. Specifically, when training the I2A model, we begin with only the image-to-notation (OMR) examples in the batch mixture; after 15k training steps, we start including the MIDI-to-audio synthesis examples; and after 50k steps, we begin adding the direct image-to-audio samples. Likewise, for the A2I model, we start with only audio-to-MIDI transcription, then add notation-to-image (score rendering) from 40k steps onward, and finally include audio-to-image examples after 70k steps. In this way, the model first learns easier or more data-rich subtasks before gradually facing the full complexity of the outer-modality translation tasks. Throughout training, we apply weighted sampling of the task datasets to balance their contributions, ensuring that no single task with a large dataset dominates the learning schedule. This careful training setup was crucial for stabilizing the joint training and achieving good performance across all translation tasks.

\begin{table*}[t]
    \caption{%
Transformer model configurations and task introduction steps.  
The upper block (first five rows) contains models for the
\emph{Image-to-Audio} direction, whereas the lower block (last four rows) contains models for the \emph{Audio-to-Image} direction.  \\
In the three ``Task Introduction'' columns, pre-slash tasks are for I2A, post-slash for A2I. Numbers indicate the training step at which each task dataset is first sampled ($0$ = from the start, "--" = not used).\\
M2A denotes MIDI-to-Audio synthesis and L2I denotes LMX-to-Image (score rendering).}

    \label{tab:model-config}
    \centering
    \renewcommand{\arraystretch}{1.15}
    \begin{tabular}{l||c|c|c|c||c|c|c}
        \hline
        \multirow{2}{*}{Models} &
        \multicolumn{4}{c||}{Model Size} &
        \multicolumn{3}{c}{Task Introduction (trainging step)} \\
        \cline{2-8}
        & Dimension & Enc/Dec Layers & Heads & Sub-Dec Heads & OMR/AMT & M2A/L2I & I2A/A2I \\
        \hline
        OMR Only                                          & 512  & 12 &  8 &  8 & 0 & --     & --     \\
        Image-to-Audio Only                               & 768  & 12 & 10 & 10 & --& --     & 0      \\
        MIDI-to-Audio Only                                & 512  &  4 &  8 &  8 & --& 0      & --     \\
        OMR + Image-to-Audio                              & 1024 & 12 & 16 &  8 & 0 & 15,000 & --     \\
        OMR + Image-to-Audio + MIDI-to-Audio              & 1024 & 12 & 16 &  8 & 0 & 15,000 & 50,000\\
        \hline
        AMT Only                                          & 768  & 12 & 12 &  8 & 0 & --     & --     \\
        Audio-to-Image Only                               & 768  & 12 & 10 & 10 & -- & --    & 0      \\
        AMT + Audio-to-Image                              & 1024 & 12 & 16 &  8 & 0 & 40,000 & --     \\
        AMT + Audio-to-Image + LMX-to-Image               & 1024 & 12 & 16 &  8 & 0 & 40,000 & 70,000\\
        \hline
    \end{tabular}
\end{table*}

\subsection{Data Split and Test Sets}
As most of our datasets focus on Western classical music repertoire, there exists a clear overlap of pieces featured between datasets; we carefully modify the assignments to avoid any overlap between splits. Among the YTSV dataset, pieces included in the test set of MAESTRO or MusicNet were assigned to the test set.
We also make sure that multiple versions of a piece in YTSV are assigned to the same split. 

The Beethoven Piano Sonata Dataset (BPSD) is the only subset in our collection that contains all four modalities, and we further preprocess a portion of it to create a test set for all translation tasks. Among the 32 sonatas in BPSD, we select 9 that are included in MAESTRO test split, and also exclude them from the training set of the GrandStaff dataset. After isolating the individual systems from the scanned images, we align each system with the corresponding sections of MusicXML, MIDI, and audio. 

To evaluate I2A results on scanned scores of various quality, we manually select a subset of 11 pieces by 7 composers from the classical era to the contemporary (named \textit{YTSV-T11}). We also manually check for duplications in the rest of YTSV.

\subsection{Evaluation Metrics}
For image-to-audio generation, we evaluate the generated audio at the MIDI level using the following approach. Given the reference and corresponding generated audio, we first apply the Onsets and Frames \cite{hawthorne2017onsets} model for piano transcription to obtain transcribed MIDI data. We use note onset, offset, and pitch information from the transcribed MIDIs to produce a time alignment between the reference and generated audio using dynamic time warping (DTW), allowing us to produce time-aligned MIDI representations for both.(Appendix \ref{appendix:dtw}) Finally, we compute the note onset \(F{_1}\) score of the time-aligned reference and generated MIDI representations using the transcription evaluation metrics from \texttt{mir\_eval}. To account for potential DTW timing misalignments, we evaluated our results using three onset tolerance thresholds (50ms, 100ms, 200ms). We also report the Frechet Audio Distance (FAD)~\cite{fadtk} between the generated results and the test set ground truths using the LAION-CLAP embedding~\cite{laionclap2023}. For MIDI-to-audio synthesis we adopt an analogous protocol but omit the dynamic time-warping step: the reference MIDI is already precisely aligned to the generated audio, so we directly compute the note-onset \(F_{1}\) score between the reference MIDI and the MIDI transcribed from the synthesis, and report the corresponding FAD under the same LAION-CLAP embedding.

For audio-to-image generation evaluation, we apply our reproduction of the Zeus~\cite{mayerZeus} OMR model to extract LMX tokens from the audio-conditioned score image generations. We then compute the Earth Mover's Distance (EMD)~\cite{Rubner2000} metric between histograms of the extracted LMX tokens and the ground truth LMX. Specifically, we construct token frequency distributions where each bin position represents a unique token and bin height represents its frequency. To provide more granular analysis of our model's performance in capturing different musical aspects, we calculate EMD separately for pitch tokens (e.g., \texttt{C4}, \texttt{G3}, with 53 distinct pitch values) and duration tokens (e.g., \texttt{half}, \texttt{quarter}, \texttt{eighth}, with 13 distinct duration values). For duration tokens, which follow a 2:1 ratio between adjacent values (e.g., a half note is twice as long as a quarter note), we account for potential meter interpretation differences in our A2I model's output (such as single 4/4 measure interpreted as two measures with doubled note duration values). We address this by computing EMD with position shifts of -1, 0, and 1 on the predicted duration distribution and selecting the minimum EMD value among these shifts, allowing for more flexible evaluation of temporal aspects. Details are elaborated in Appendix \ref{appendix:emd}.

For OMR evaluation, we use the Symbol Error Rate (SER) metric on LMX token sequences, following \cite{mayerZeus}. For AMT evaluation, we use Note-\(F_1\) score implemented in the \texttt{mir\_eval} \cite{raffel2014mir_eval} library. As our main focus is western classical music, we exclude results for SLakh. For MusicNet, we use the string and wind test split following previous works~\cite{cheuk2021reconvat, maman22a, chang2024yourmt3+}.


\section{Results}

\subsection{Image-to-Audio Generation}

\begin{table*}[ht]
    \centering
    \caption{%
Image‐to‐audio accuracy reported as onset $F_{1}$ (↑) and FAD (↓). 
Note that the two rows marked \emph{OMR + Image-to-Audio + MIDI-to-Audio} share a
\textbf{single} model jointly trained on those three tasks.
“Direct I2A” feeds a score image to the model and predicts audio in one
shot, whereas “Multi-stage” first applies the model for OMR (image → MusicXML), converts that MusicXML to MIDI with a simple rule-based conversion with fixed tempo and MIDI, and then reuses the same model for M2A (MIDI → audio).
As YTSV-T11 consists of multi-system images (i.e. having two or more systems per image), we did not report multi-stage results as the model was not trained for multi-system OMR.}
    \label{tab:i2a-result}
    \begin{small}
    \begin{tabular}{l|c||ccc|ccc|c|c}
        \hline
        & Metric & \multicolumn{6}{c|}{$F_1$ Score $\uparrow$} & \multicolumn{2}{c}{FAD $\downarrow$} \\
        \cline{2-10}
        Method & Dataset & \multicolumn{3}{c|}{BPSD} & \multicolumn{3}{c|}{YTSV-T11} & BPSD & YTSV-T11 \\
        \cline{2-10}
        & Onset Tolerance (ms) & 50 & 100 & 200 & 50 & 100 & 200 &  -- & --  \\
        \hline
        \multicolumn{2}{l||}{Direct I2A: YTSV-P (Image-to-Audio Only)} & 23.49 & 34.51 & 44.15 & 27.05 & 43.32 & 53.02 & 0.422 & 0.317 \\
        \multicolumn{2}{l||}{Direct I2A: OMR + Image-to-Audio} & 48.67 & 64.30 & 74.01 & 51.60 & 67.92 & 75.98 & 0.098 & 0.056 \\
        \multicolumn{2}{l||}{Direct I2A: OMR + Image-to-Audio + MIDI-to-Audio} & 48.36 & 64.63 & 74.92 & 52.66 & \textbf{68.45} & \textbf{76.24} & \textbf{0.081} & \textbf{0.055} \\ 
        \hline
        \multicolumn{2}{l||}{Multi-stage: OMR + Image-to-Audio + MIDI-to-Audio} & \textbf{50.91} & \textbf{70.40} & \textbf{79.96} & -- & -- & -- & 0.137 & -- \\
        \multicolumn{2}{l||}{Multi-stage: Zeus → VirtuosoNet → MSD} & 45.52 & 59.35 & 69.36 & -- & -- & -- & 0.330 & -- \\ 
        \hline
        \multicolumn{2}{l||}{DAC Reconstruction (Upper-bound)} & 68.83 & 82.39 & 87.47 & 82.28 & 86.43 & 88.76 & 0.050 & 0.035 \\
        \hline
    \end{tabular}
    \end{small}
\end{table*}

Generating music audio directly from a given score image has never been explored in an end-to-end manner. Our result in \cref{tab:i2a-result} shows that our I2A model trained with multiple tasks successfully generates performance audio that corresponds to the input score image, showing high $F_{1}$ score. To demonstrate the performance bottleneck from audio tokenization, we also measure the metric for DAC-reconstructed audio of reference performance, which can be regarded as the upper limit.

As mentioned earlier, the model only trained for image-to-audio task could not generate audio that is coherent to input score image as shown by low transcription result nor fluent audio as shown by high FAD. On the other hand, the model trained along with the OMR task showed significant improvement in both transcription score and FAD, showing clear evidence that the model read the notes from input score image to generate corresponding audio. Adding MIDI-to-audio task improved note reconstruction accuracy at YTSV test set. The model combined with the MIDI-to-audio task showed an improved FAD score, which indicates that the model learned modeling of more natural sound. The result shows encompassing these different modal translation tasks improved the model's performance.

To assess the effectiveness of our proposed end-to-end approach without explicit symbolic intermediate,
we also run our unified model in a \emph{multi-stage procedural cascade}: the image is first passed through the model in OMR mode (Image\(\rightarrow\)MusicXML), the resulting MusicXML is converted to MIDI, and the same model is then invoked in its MIDI\(\rightarrow\)Audio mode. For external comparison, we also experimented a specialist baseline comprising Zeus for OMR~\cite{mayerZeus}, VirtuosoNet for performance modelling~\cite{jeong2019graph}, and Music-Spectrogram Diffusion (MSD) for audio synthesis~\cite{multiinst}.

When comparing the multi-stage approach to the direct I2A method on the same model(trained on OMR + I2A + Midi-to-Audio), its $F_1$ scores were better but the corresponding FAD value was worse. In practice, this translates to more accurate note timing and pitch on paper, yet a noticeably lower perceptual quality—often manifesting as dissonant or “off-sounding” passages. The bottleneck mainly comes from the OMR errors. We found that errors in the OMR stage often result in perceptually obvious flaws, such as strong dissonance. This stems from a fundamental limitation of current OMR models: they are trained to optimize visual accuracy without any awareness of the downstream auditory impact. However, visually subtle errors, such as misreading an accidental or misclassifying a note by a semitone, can lead to musically severe consequences in the audio output. Since the OMR trakining lacks any feedback from how its predictions will ultimately sound, it cannot learn to avoid such acoustically significant mistakes.

\begin{figure}[t]
    \begin{center}
    \includegraphics[width=1.0\columnwidth]{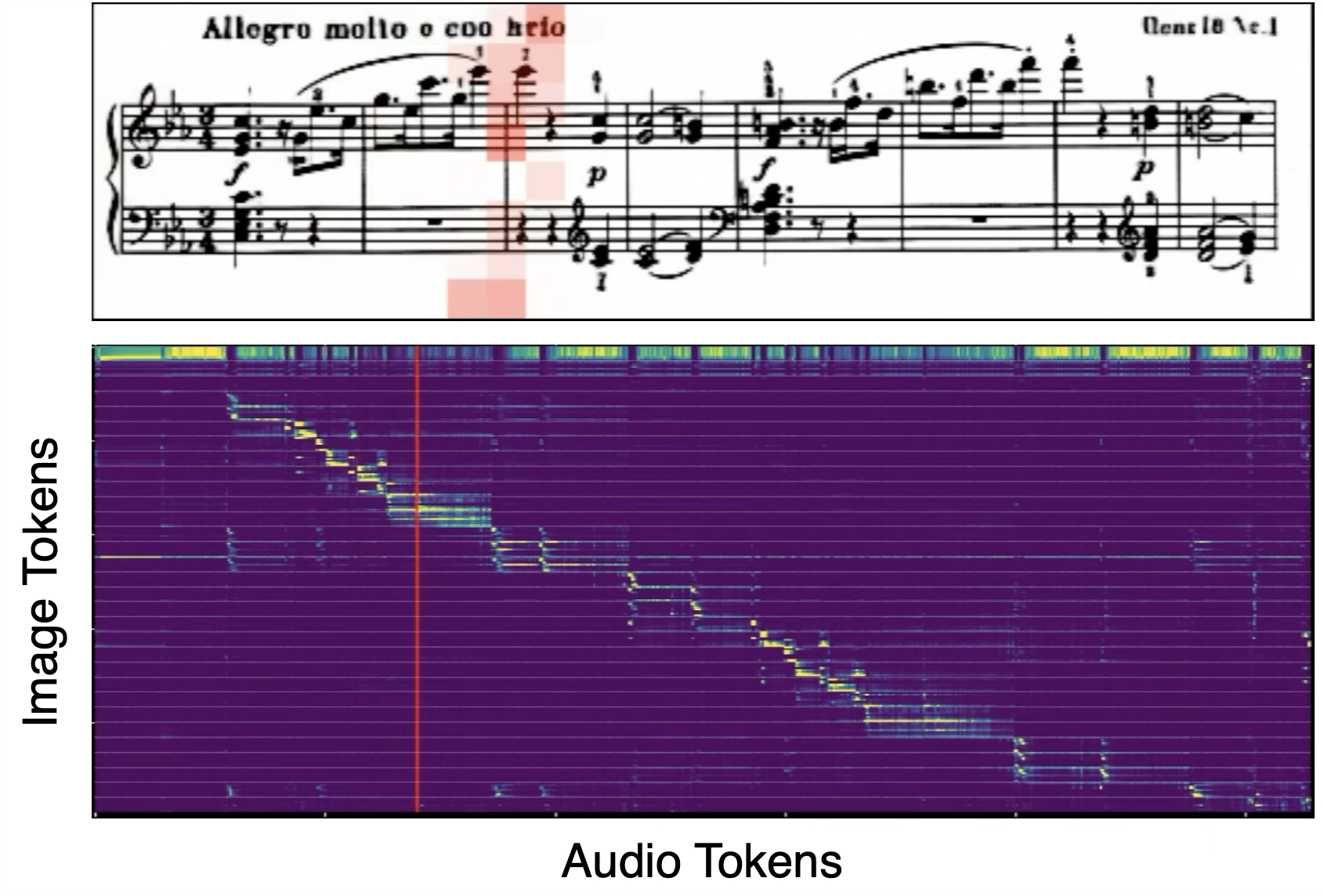}
    \caption{Attention patterns from a selected transformer head. Top: Red heatmap shows attention weights across RVQ token regions in the image. Bottom: Token-level attention matrix with audio tokens (x-axis) versus image tokens (y-axis), revealing cross-modal relationships.}
    \label{fig:attention_vis}
    \end{center}
\end{figure}

To validate that the model actually attends to the corresponding parts of score image to generate performance audio, we visualized the attention map of the model as shown in Fig.~\ref{fig:attention_vis}. The result clearly shows that the model followed the score to translate it to performance audio.

To evaluate the quality of the generated results, we strongly encourage readers to refer to the actual audio and video examples provided on our demo page\footnote{\url{https://sakem.in/u-must/}}.

\subsection{Audio-to-Image Generation}

\begin{table}[t]
    \vskip -0.1in
    \centering
    \caption{Audio-to-image generation accuracy in EMD on BPSD.}
    \vskip 0.1in
    \begin{small}
    \begin{tabular}{l||cc}
        \hline
        \multirow{2}{*}{Method} & \multicolumn{2}{c}{EMD $\downarrow$} \\
        \cline{2-3}
        & Pitch & Duration \\
        \hline
        Audio-to-Image Only & 4.6436 & 0.4873 \\
        + AMT & 2.8880 & 0.4377 \\
        + LMX-to-Image & \textbf{2.6350} & \textbf{0.4317} \\
        \hline
        GT Random Pairing Baseline & 3.4921 & 0.9936 \\
        \hline
        RQVAE Reconstruction & 0.8990 & 0.1301 \\
        GT Image & 0.4865 & 0.1113 \\
        \hline
    \end{tabular}
    \end{small}
    \label{tab:a2i}
\end{table}

Converting audio recordings into sheet music requires navigating the complex transformation from acoustic signals to structured visual notation. Our results in \cref{tab:a2i} highlight the progress of our A2I model in this challenging domain. To contextualize our evaluation metrics, we include measurements for images reconstructed using the RQVAE, ground truth images processed through our OMR system, and a ground truth random pairing baseline--where we compute the EMD between one ground truth sample and the remaining ground truth samples--as reference points. Interestingly, the ground truth variability baseline indicates better performance than the A2I model trained without additional tasks, underscoring the difficulty of the A2I task without further guidance. The integration of auxiliary tasks proves beneficial for A2I performance. Our model, trained with additional AMT and LMX-to-image tasks, indicates improved EMD metrics compared to the baseline trained solely on A2I, suggesting that explicitly modeling intermediate symbolic representations enhances the model's ability to generate appropriate musical notation.

The consistently higher pitch EMD values compared to duration EMD values observed in the table reflect the structural difference in token space sizes rather than indicating lower pitch accuracy. With 53 distinct pitch tokens versus only 13 duration tokens, the Earth Mover's Distance naturally yields larger values when measuring transportation distances across the wider pitch space, making direct comparison between these metrics less meaningful than comparing relative performance within each category.

\begin{figure}[t!]
    \begin{center}
    \includegraphics[width=0.8\columnwidth]{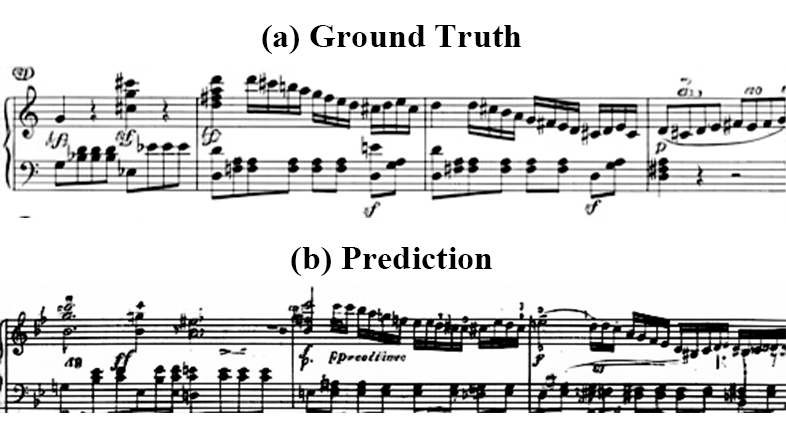}
    \caption{One example from audio-to-image translation.}
    \label{figure-a2i-output}
    \end{center}
\end{figure}

Although the generated result does not satisfy human-readable and performable quality standards, the result shows that the A2I model partially reflects the notes played in the audio, as shwon in Fig.~\ref{figure-a2i-output}. The asymmetry in generation quality between I2A and A2I might come from the intrinsic difficulty of generating systematic notation from audio, which was also discussed in previous literature~\cite{galan2024muscat}. Also, the fact that our training image-audio pairs include artificial concatenation between multiple systems could have introduced additional challenges.

\subsection{MIDI-to-Audio Synthesis}

\begin{table}[t]
    \centering
    \caption{MIDI-to-audio synthesis accuracy in \(F_1\) and FAD on BPSD.}
    \begin{small}
    \begin{tabular}{l||ccc|c}
        \hline
        \multirow{2}{*}{Method} & \multicolumn{3}{c|}{\(F_1\) $\uparrow$} & \multirow{2}{*}{FAD $\downarrow$} \\
        \cline{2-4}
        & 50ms & 100ms & 200ms & \\
        \hline
        MIDI-to-Audio Only & 26.61 & 64.86 & \textbf{88.20} & 0.201 \\
        + OMR + I2A & \textbf{39.37} & \textbf{66.63} & 84.66 & \textbf{0.143} \\
        \hline
    \end{tabular}
    \end{small}
    \label{tab:m2a}
\end{table}

Table~\ref{tab:m2a} summarizes MIDI-to-Audio (M2A) performance on the Beethoven Piano Sonata Dataset. 
A model trained solely on M2A achieves onset \(F_1\) scores of 26.61, 64.86, and 88.20 in 50ms, 100ms, and 200ms tolerance windows, respectively, with a Fréchet Audio Distance (FAD) of 0.201. Augmenting training with OMR and Image-to-Audio tasks yields pronounced gains at stricter tolerances: the 50ms \(F_1\) rises to 45.91 (+19.3 pp) and the 100 ms score to 69.54 (+4.7 pp), while the 200ms result remains comparable (84.55). Concomitantly, FAD decreases by 42 \% to 0.116, indicating markedly improved perceptual quality. These findings confirm that incorporating complementary score-conditioned tasks sharpens temporal precision and enhances audio realism in M2A synthesis.

\subsection{OMR and AMT}
To validate the effectiveness of incorporating image-to-audio and audio-to-image for solving other tasks, we also train models using only OMR datasets (GrandStaff and Olimpic) or AMT datasets (MAESTRO, MusicNet, and SLakh). As the size of the datasets differs greatly from YTSV, we experiment with various model sizes and select the one with the best performance on the validation set. The selected baseline OMR-only model consists of an encoder and a decoder, each with 12 layers, 512 dimensions, and 8 attention heads, and the AMT-only model consists of the same number of layers but with 768 dimensions and 12 attention heads.

\begin{table}[t]
\centering
\caption{OMR Results in SER. Lower is better.}
\begin{small}
\begin{tabular}{l || r r | r}
\hline
\multirow{2}{*}{Method} & \multicolumn{2}{c|}{OLiMPiC} & BPSD \\ 
\cline{2-4}
& Synth & Scanned & Scanned \\ 
\hline
OMR-only         & 15.90  & 24.58  & 45.39 \\ 
+ Image-to-Audio & 10.57  & 15.45 &  23.85 \\
+ MIDI-to-Audio  & \textbf{9.72} & \textbf{13.67}  & \textbf{23.36} \\ 
\hline
Zeus             &  10.10  & 14.45  & 31.24 \\ 
\hline
\end{tabular}
\end{small}
\label{tab:omr_result}
\end{table}

\cref{tab:omr_result} provides OMR evaluation results on OLiMPiC and BPSD-test. As the original paper of Zeus~\cite{mayerZeus}, the state-of-the-art model in piano-form OMR, does not include results of a model trained on both OLiMPiC and GrandStaff, we train a reproduction model for comparison with the same data split we used.
The results show that performance increases as more modal translation tasks were added, and our unified approach achieved the best results.
The addition of MIDI-to-audio translation, a task that has no modal overlap with OMR, also improved the performance of OMR. We hypothesize that this is because the model is trained to follow note-level conditions in MIDI-to-audio, which guides the model to better handle similar information in score image inputs.

\begin{table}[t]
\centering
\caption{AMT results in note onset \(F_1\) score for test set, compared to Maman \emph{et al.} \cite{maman22a}. Higher is better.}
\begin{small}
\begin{tabular}{l || r r | r}
\hline
\multirow{2}{*}{Method} & \multicolumn{2}{c|}{MusicNet} & \multirow{2}{*}{MAESTRO} \\ 
\cline{2-3}
& Str & WW & \\ 
\hline
AMT-only          & 87.21  & 72.04  & 89.40 \\ 
+ Audio-to-Image  & \textbf{87.28}  & 72.61  & 89.38 \\ 
+ LMX-to-Image    & 87.25  & \textbf{75.52}  & \textbf{89.45} \\ \hline
Maman \emph{et al.}      & 81.8   & 84.2   & 89.7 \\ 
\hline
\end{tabular}
\end{small}
\label{tab:amt_result}
\end{table}

\cref{tab:amt_result} presents the results of AMT in terms of Note-\(F_1\) scores. Unlike in OMR, incorporating audio-to-image translation does not lead to noticeable improvements. We attribute this to the already extensive training data available for AMT such as MAESTRO, which provides 200 hours of piano audio with precisely aligned MIDI annotations.
We also compare our model to \cite{maman22a}, which used additional 73 hours of unaligned score and audio for training. Our model performs better for string instruments even without additional training, while performing worse in woodwind. We believe the difference comes from dataset distribution. While the dataset used by Maman \emph{et al.} included 26 hours of orchestral recordings, our dataset did not, thus heavily lacking woodwind sounds; on the other hand, the higher proportion of string instrument samples in our dataset may have provided our representation model with a learning advantage in capturing the distinct timbre characteristics.


\section{Conclusion}
In this work, we presented the first unified music modal translation model capable of successful score image-to-audio generation, along with state-of-the-art performance in optical music recognition. We believe that this result highlights the potential of integrating various modalities for a more holistic understanding and processing of music information.

We also presented the YouTube Score Video (YTSV) dataset, a large-scale collection of score image and performance audio pairs. In particular, our experiment results with YTSV emphasize the importance of scanned music scores as a valuable source of musical data, an area that has received relatively less attention in music information retrieval compared to the audio and symbolic domains. Given the vast number of publicly available scanned scores in the International Music Score Library Project (IMSLP), we expect that incorporating these resources could have a significant impact on music understanding and generation.

However, there are clear limitations for the current study.  
First, the improvement in AMT accuracy is modest relative to the substantial gains observed for OMR and I2A translation.  
Second, the quality of the A2I outputs remains insufficient for practical use: rendered pages frequently exhibit overcrowded staves, misplaced clefs, and other notational artifacts.
Also, the framework currently employs two independent checkpoints for the I2A and A2I directions. Addressing this issue and training a single unified model is remained for future research. While the current framework employed the encoder-decoder structure, a decoder-only model can be considered as a powerful tool for unifying all the modalities in a single model.

Looking forward, our YTSV dataset opens several promising research directions. We plan to explore self-supervised learning for more precise image-audio alignment, develop score-conditioned audio tasks like expressive performance modeling, and investigate methods to extend sparse slide-level alignments to note-level supervision. Additionally, improving the underlying audio and image codecs remains an important goal to enhance both the visual quality of our A2I outputs and the acoustic fidelity of I2A generation.

\bibliographystyle{ieeetr}  
\bibliography{refs}   

{\appendices

\section{Mathematical Formulation of Multimodal Tokenization and Unified Translation Model}\label{sec:math_formulation}
\subsection{Tokenizers and Unified Vocabulary}

We denote a \emph{modality} by the calligraphic symbol
\(
\mathcal{X}\!\in\!\{\mathcal{I},\mathcal{A},\mathcal{N},\mathcal{M}\}
\)
for \textbf{I}mage, \textbf{A}udio, musical \textbf{N}otation (LMX), and \textbf{M}IDI performance, respectively.  
The corresponding \emph{raw} data \(X\) are written with roman capitals
\(I, A, N, M\).  
Each modality owns a \emph{tokenizer encoder}
\(
\mathcal{F}_{\mathcal{X}}
\)
and an inverse \emph{tokenizer decoder}
\(
\mathcal{G}_{\mathcal{X}}
\)
such that
\begin{equation}
\mathcal{F}_{\mathcal{X}}(X)=
z^{(\mathcal{X})}_{1:L_{\!X}},
\qquad
\mathcal{G}_{\mathcal{X}}\!\bigl(z^{(\mathcal{X})}_{1:L_{\!X}}\bigr)\approx X .
\end{equation}
Hence, \(z^{(\mathcal{X})}_{1:L_{\!X}}\) is the discrete-token representation of \(X\) and \(L_{\!X}\) is its length (we subsequently write \(L_{I},L_{A},L_{N},L_{M}\) for the four modalities).  
Every modality has its own vocabulary \(\mathcal{V}_{\!\mathcal{X}}\), yet all tokens ultimately live in a \emph{shared} space \(\mathcal{V}\) (Sec.~\ref{sec:tokenization:unified_vocab}), allowing a single Transformer to translate between any pair of modalities.

For the \textbf{continuous} modalities—score images and audio—we learn residual vector-quantised tokenizers. Both tokenizers employ \(d=4\) \emph{unshared} codebooks, each of cardinality
\(
\kappa = 1024
\).
Consequently, each time-step yields a \emph{bundle} of \(d\) code indices
\(z_{t,1},\dots,z_{t,d}\in\{0,\dots,\kappa\!-\!1\}\),
so the discrete representation is a 2-D array of shape
\(L_{\!X}\!\times\! d\).
We still refer to its “length’’ as \(L_{\!X}\) for notational simplicity.

\paragraph{Score images, \(\mathcal{X}=\mathcal I\)}

RQVAE compresses each image patch by a factor of \(C\!=\!16\).
Given a score image that contains \(K\) musical systems
\(
\text{system}_{i}\in\mathbb{R}^{W_{i}\times H_{i}}
\),
RQVAE produces
    
\begin{equation}
\text{system}_{i,\text{tokens}}
       \;\in\;
       \Bigl(\mathcal{V}_{\!\mathcal{I}}\Bigr)^{\bigl(\frac{W_{i}H_{i}}{C^{2}}\times d\bigr)}.
\end{equation}

We flatten each \(d\!\times\!\frac{W_{i}H_{i}}{C^{2}}\) token grid in \emph{vertical} reading order (top-to-bottom inside a column, then left-to-right between columns) and concatenate all systems, inserting the separator token \texttt{[SEP]}:

\begin{multline}
\text{image}_{\text{tokens}} =
\text{Concat}\bigl(\text{system}_{1,\text{tokens}},\; \texttt{[SEP]},\; \dotsc,\; \\
\texttt{[SEP]},\; \text{system}_{K,\text{tokens}}\bigr).
\end{multline}

\begin{equation}
\left| \text{system}_{i,\text{tokens}} \right| = \frac{W_i \times H_i}{C^2} \, d, \\[5pt]
\end{equation}

\begin{equation}
L_{I} = \sum_{i=1}^K \left| \text{system}_{i,\text{tokens}} \right| + (K-1)
\end{equation}

\paragraph{Audio, \(\mathcal{X}=\mathcal A\)}
All audio is resampled to \(f_s=44.1\,\text{kHz}\) (mono) and tokenized with DAC using hop size \(h=512\) samples.  
The number of hop frames in \(T\) seconds equals
\( \bigl\lceil\tfrac{T f_s}{h}\bigr\rceil\), giving

\begin{equation}
L_{A}= \Bigl\lceil \tfrac{T f_s}{h} \Bigr\rceil,
\qquad
\text{audio}_{\text{tokens}}
      \;\in\;
      \bigl(\mathcal{V}_{\!\mathcal{A}}\bigr)^{L_{A}\times d}.
\end{equation}

\paragraph{Linearized MusicXML, \(\mathcal{X}=\mathcal N\)}
Notation data are stored as linearized MusicXML(LMX)~\cite{mayerZeus}.
Let \(L_{N}\) be the sequence length; the token string is

\begin{equation}
\text{lmx}_{\text{tokens}}
      \;\in\;
      \bigl(\mathcal{V}_{\!\mathcal{N}}\bigr)^{L_{N}\times 1},
\end{equation}

where by convention \(d=1\).  
For training and inference we \emph{pad} codebook positions 2–4 with a special \texttt{PAD} token:
\(z_{t,2:4}^{(\mathcal{N})}=\texttt{PAD}\).

\paragraph{MIDI, \(\mathcal{X}=\mathcal M\)}
We adopt the \textsc{YourMT3+}~\cite{chang2024yourmt3+} MIDI-like event vocabulary (10ms quantisation).  
If \(L_{M}\) denotes its length,

\begin{equation}
\text{midi}_{\text{tokens}}
      \;\in\;
      \bigl(\mathcal{V}_{\!\mathcal{M}}\bigr)^{L_{M}\times 1},
\end{equation}
and again \(z_{t,2:4}^{(\mathcal{M})}=\texttt{PAD}\).

\paragraph{Unified Vocabulary, $\mathcal{V}$}\label{sec:tokenization:unified_vocab}

\begin{equation}
\mathcal{V}
=
\Bigl(\!\bigcup_{\mathcal{X}}\mathcal{V}_{\!\mathcal{X}}\!\Bigr)
\cup
\Bigl(\!\bigcup_{\mathcal{X}}\{\texttt{[SOS]}_{\mathcal{X}},
                                \texttt{[EOS]}_{\mathcal{X}}\}\Bigr)
\cup
\{\texttt{[SEP]},\texttt{[PAD]}\}.
\end{equation}

\subsection{Model Architecture}

\paragraph{Input embedding}
For source modality \(\mathcal{X}\) and target modality \(\mathcal{Y}\),
each input token \(z_i\) is embedded as
\begin{equation}
e_i=\underbrace{\text{TokEmb}(z^{(\mathcal{X})}_i)}_{\text{token}}
    +\underbrace{\text{PosEmb}_{\mathcal{X}}(i)}_{\text{modality-specific pos.~enc.}}
    +\underbrace{\text{TgtEmb}_{\mathcal{Y}}}_{\text{target hint}}.
\end{equation}

\paragraph{Sequence-to-Sequence Model} The encoder \(\mathcal{E}\) takes as input the source token sequence $z^{(\mathcal{X})}_{1:L_{\!X}}$ (of length $L_{\!X}$) and maps it to a sequence of hidden representations; the decoder \(\mathcal{D}\) autoregressively generates the target sequence $\hat{z}^{(\mathcal{Y})}_{1:L_{\!Y}}$ token by token.

\begin{multline}
z^{(\mathcal{X})}_{1:L_{\!X}}
     =\mathcal{F}_{\mathcal{X}}(X),\qquad
H                                  =\mathcal{E}\!\bigl(z^{(\mathcal{X})}_{1:L_{\!X}}\bigr),\qquad
\hat{z}^{(\mathcal{Y})}_{1:L_{\!Y}}=\mathcal{D}(H),
\end{multline}
and finally
\(
\hat{Y}=\mathcal{G}_{\mathcal{Y}}\!\bigl(\hat{z}^{(\mathcal{Y})}_{1:L_{\!Y}}\bigr)
\).

\paragraph{Sub-Decoder Module $\mathcal{D}_{sub}$}

The main decoder \(\mathcal{D}_{main}\) emits a hidden state \(h_t\in\mathbb{R}^{D}\)
for each \emph{time step} \(t\).
To transform \(h_t\) into \(d\) code tokens we introduce a one-layer Transformer 
\emph{sub-decoder} \(\mathcal{D}_{sub}\):

\begin{equation}
   (z_{t,1},\dots,z_{t,d})=\mathcal{D}_{sub}\bigl(h_t\bigr)
\end{equation}

\begin{itemize}
\item \textbf{Input to \(\mathcal{D}_{sub}\).}  
      For sub-step \(\ell\) (\(1\!\le\!\ell\!\le\!d\)) we sum  
      (i) the token embedding of the previous output
      \(\hat{z}_{t,\ell-1}\) (or $null$ if \(\ell=1\)),
      (ii) a dedicated positional embedding for \(\ell\),
      and (iii) the broadcast main-decoder state \(h_t\):
      \(
      e_{t,\ell}= \text{TokEmb}_{sub}(\hat{z}_{t,\ell-1})
                 +\text{PosEmb}_{sub}(\ell)
                 +h_t .
      \)

\item \textbf{Symbolic targets.}  
      If \(\mathcal{Y}\in\{\mathcal{N},\mathcal{M}\}\) we set \(d=1\);  
      codebooks \(2\!-\!4\) are always \texttt{PAD}.
\end{itemize}

\subsection{Training Objective}

With cross-entropy loss
\(
\mathrm{CE}(\cdot,\cdot)
\)
and parameter set \(\theta\),

\begin{equation}
\!\!\mathcal{L}(\theta)=
-\!\!\!\sum_{t=1}^{L_{\!Y}}\!
  \begin{cases}
    \displaystyle
    \sum_{\ell=1}^{d}
       \log P_{\theta}\!\bigl(z_{t,\ell}^{\text{(gt)}} \mid
                               z_{<t,*}^{\text{(gt)}},
                               H
                         \bigr),
       & \!\!\!\!\mathcal{Y}\in\{\mathcal{I},\mathcal{A}\} \\
    \log P_{\theta}\!\bigl(z_{t,1}^{\text{(gt)}} \mid
                            z_{<t,1}^{\text{(gt)}}, H
                      \bigr),
       & \!\!\!\!\mathcal{Y}\in\{\mathcal{N},\mathcal{M}\}.
  \end{cases}
\end{equation}

When applying $softmax$ for each modality, entries of other modalities are masked as $-inf$.

\section{Evaluation Details}

\subsection{DTW Alignment for Onset $F_1$ Computation}\label{appendix:dtw}

\begin{figure}[t]
    \subfloat[]{\includegraphics[height=0.20\linewidth]{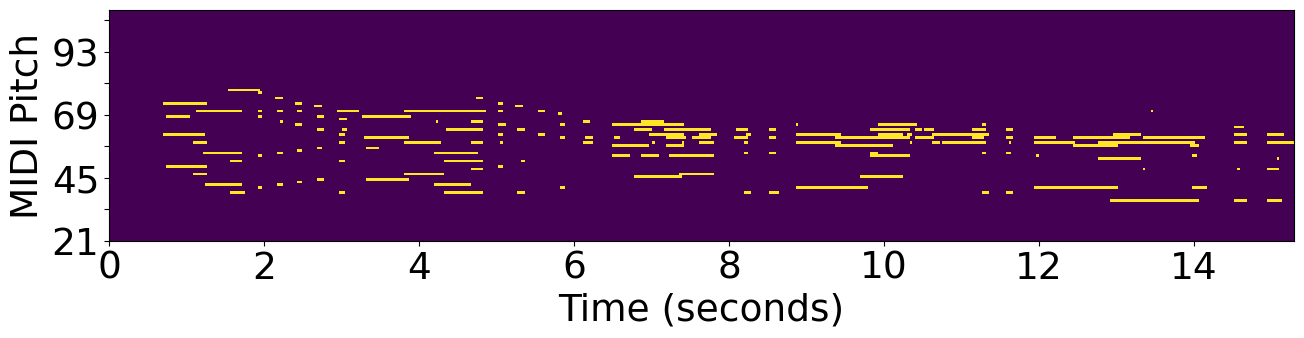}} \\
    \subfloat[]{\includegraphics[height=0.20\linewidth]{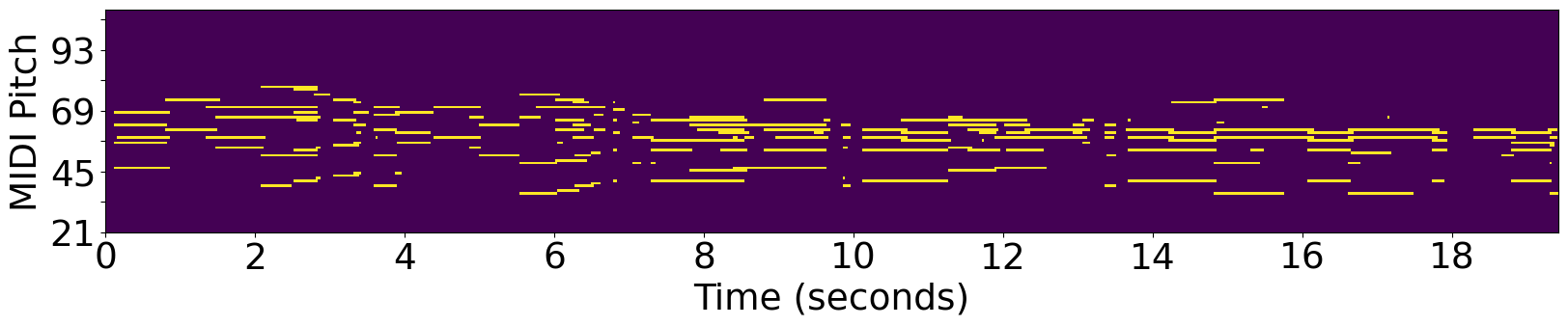}} \\
    \subfloat[]{\includegraphics[height=0.20\linewidth]{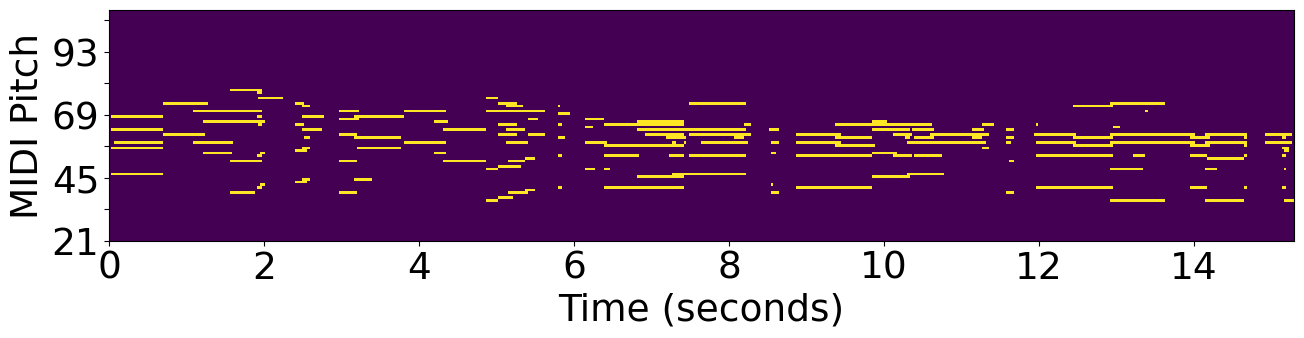}}
    \caption{Piano roll visualizations of MIDI representations from (a): Ground Truth, (b): Transcription from our Image-to-Audio generation model output, and (c): DTW-alignment of (b) on (a) for evaluation.
    }
    \label{fig:dtw}
\end{figure}

To evaluate onset detection accuracy while accounting for potential temporal misalignments, we employ Dynamic Time Warping (DTW) alignment between the ground truth frames and frames transcribed from our I2A model-generated audio. The alignment operates in one dimension by applying temporal warping across the entire piano roll along the time axis. This approach allows us to handle variations in tempo and timing while preserving the pitch information. The resulting alignments enable more accurate computation of onset $F_1$ scores by pairing the corresponding events despite temporal differences. \cref{fig:dtw} illustrates the alignment process and demonstrates how DTW effectively matches onset events between ground truth and generated sequences.

\subsection{Token Distribution Histograms for EMD Computation}\label{appendix:emd}

\begin{figure}[t]
    \centering
    \subfloat[]{\includegraphics[width=1.0\columnwidth]{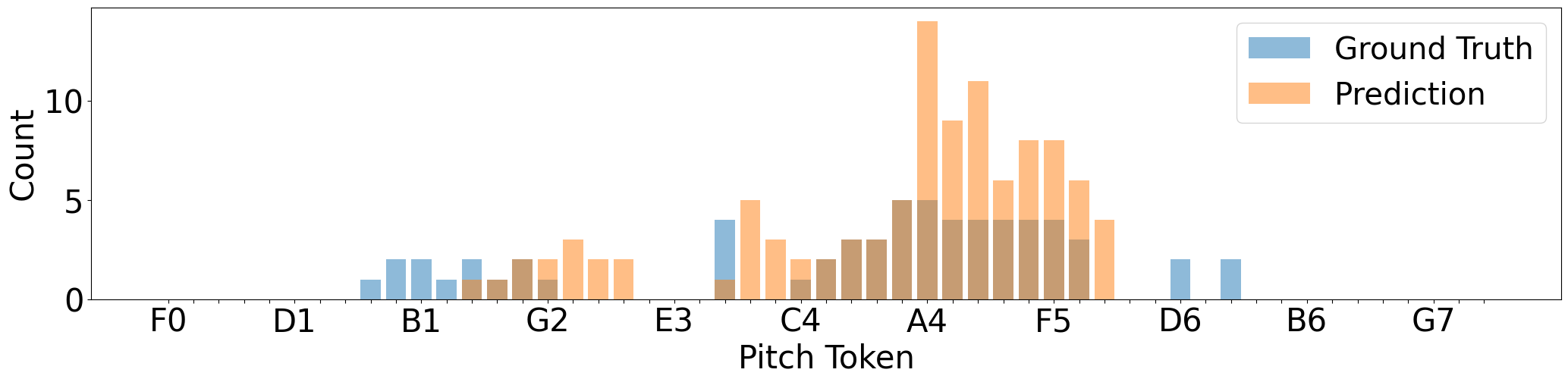}}\\
    \subfloat[]{\includegraphics[width=0.48\columnwidth]{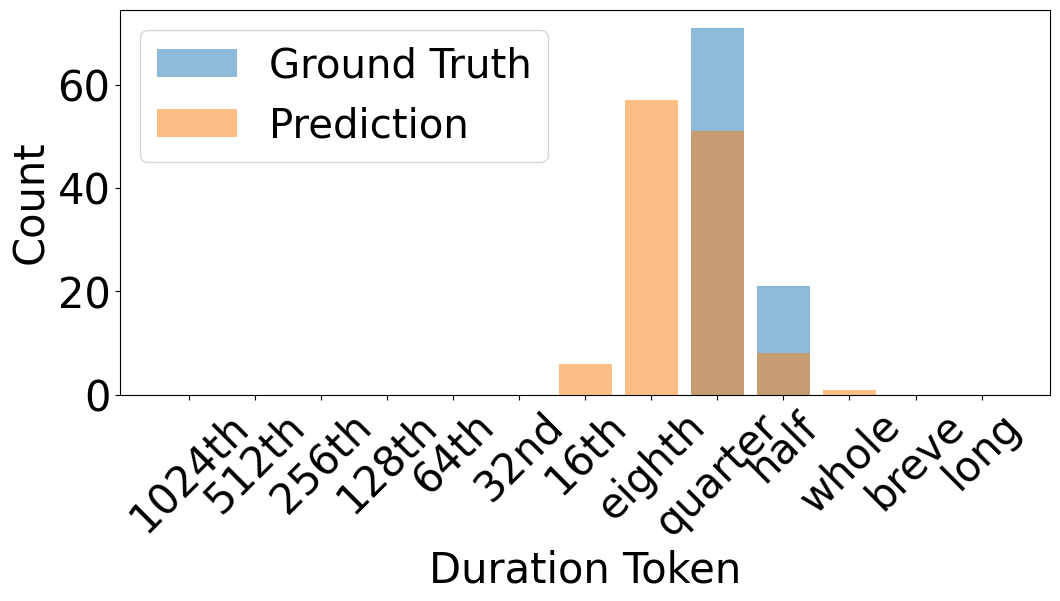}} \quad
    \subfloat[]{\includegraphics[width=0.48\columnwidth]{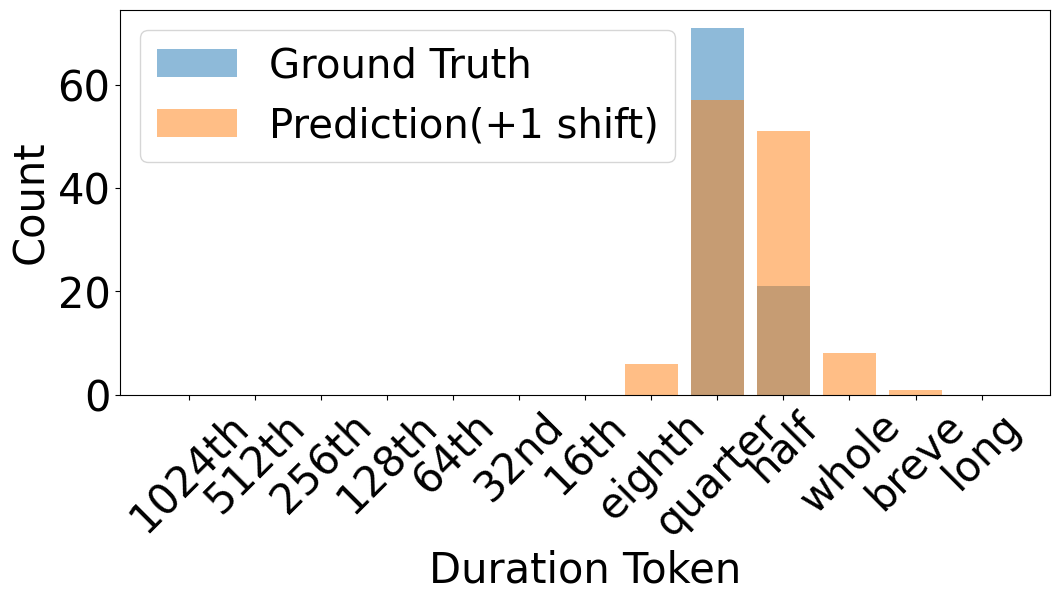}\label{fig:shifted_emd_hist}}
    \caption{Token distribution histograms used for Earth Mover's Distance (EMD) calculation. (a): Pitch token distribution histogram, (b): Duration token distribution histogram (no shift), (c): Duration token distribution histogram with +1 shift applied to prediction. The histograms show the frequency of each token type in ground truth versus prediction. For duration tokens, applying a +1 shift improves alignment between distributions, indicating our model may have interpreted the meter differently than the ground truth.}
    \label{fig:emd_histograms}
\end{figure}

In this section, we present histograms of token distributions used for Earth Mover's Distance (EMD) computation on A2I models. \cref{fig:emd_histograms} shows the token distributions for (a) pitch tokens, (b) duration tokens without shift, and (c) duration tokens with +1 shift applied. These visualizations demonstrate how the distributions of ground truth and predicted tokens compare across different musical dimensions, and how shifting affects the alignment between distributions for duration tokens.

The histograms illustrate why we calculate EMD on these distributions and how shifting can improve the alignment between predicted and ground truth duration tokens. As shown in \cref{fig:shifted_emd_hist}, applying a +1 shift to the prediction creates a better overlap with the ground truth distribution, resulting in a lower EMD value and more accurate evaluation of the model's temporal performance.

\section{YouTube Score Video Dataset}
\subsection{Metadata Extraction}
To facilitate content-based filtering of our dataset, we employ the Claude-3.5-Sonnet (\texttt{claude-3-5-sonnet-20241022})~\cite{claude} model to extract structured metadata from video titles. The model processes each video title and returns a standardized set of musical and technical attributes. Here is a detailed breakdown of the extracted fields:

\begin{table*}[h]
\centering
\caption{Content Metadata Fields}
\label{tab:metadata_fields}
\begin{tabular}{p{0.15\textwidth}p{0.5\textwidth}p{0.25\textwidth}}
\hline
\textbf{Field} & \textbf{Description} & \textbf{Example Value} \\
\hline
YT Id & Unique YouTube video identifier & \texttt{0oRyPLnPeFw} \\
\hline
Title of Video & Original video title as displayed on YouTube & \texttt{Walton - Passacaglia (1982) for solo cello [w/ score]} \\
\hline
Duration & Video length in \texttt{MM:SS} format & \texttt{10:06} \\
\hline
Composer Full Name & Complete name of the composer & \texttt{William Walton} \\
\hline
Title of piece & Name of the musical composition & \texttt{Passacaglia} \\
\hline
Opus number & Catalog number of the composition (null if unavailable) & \texttt{null} \\
\hline
Instrumentation & Categorization of musical forces from predefined set:\newline \texttt{[orchestral, concerto, solo, duet, trio, quartet, quintet, larger chamber music, choral, wind band, non-classical, vocal, unknown]} & \texttt{solo} \\
\hline
Category & Specific genre or form description & \texttt{cello solo} \\
\hline
Piano Included & Boolean indicating presence of piano part & \texttt{False} \\
\hline
String Included & Boolean indicating presence of string instruments & \texttt{True} \\
\hline
Wind Included & Boolean indicating presence of wind instruments & \texttt{False} \\
\hline
Voice Included & Boolean indicating presence of vocal parts & \texttt{False} \\
\hline
Year & Year of composition & \texttt{1982} \\
\hline
Staff Count & Two numbers indicating single-melody instrument staves and piano staves, separated by hyphen & \texttt{1-0} \\
\hline
\end{tabular}
\end{table*}

The model receives a prompt containing the video ID and title, along with an example of the expected output format. It then analyzes the title to extract and structure this information, maintaining consistency with predefined categories and formats. This structured metadata enables systematic filtering of our dataset based on musical characteristics, instrumentation, and historical period.

The Staff Count field deserves particular attention as it provides crucial information about score complexity. The format "X-Y" represents X single-melody instrument staves and Y piano staves. For example, "1-0" indicates one melodic staff with no piano staves, while "0-2" would indicate a piano-only piece with the typical grandstaff(two-staff) layout.

The aggregated results of the categories from extracted metadata are shown in the \cref{tab:aggregated_metadata}.

This aggregation is derived from the subset of metadata that meets specific target criteria. The dataset includes only video data that contain valid sheet music images. It consists of compositions from before the 2000s. Additionally, videos including vocal performances, as well as compositions classified under orchestra, larger chamber music, non-classical, and unknown have been excluded from this analysis.

\begin{table*}[h]
    \centering
    \caption{Aggregated Category Counts and Durations of Metadata}
    \label{tab:aggregated_metadata}
    \begin{tabular}{p{0.18\textwidth}p{0.4\textwidth}p{0.05\textwidth}p{0.1\textwidth}p{0.1\textwidth}}
        \toprule
        \textbf{Category} & \textbf{Description} & \textbf{Videos} & \textbf{Segments} & \textbf{Duration (hrs)} \\
        \midrule
        Piano Solo & Solo piano compositions & 9052 & 232029 & 762.34 \\
        \hline
        Accompanied Solo & Solo compositions for a non-piano instrument with piano accompaniment & 912 & 47373 & 141.83 \\
        \hline
        String Quartet & Compositions for two violins, viola, and cello & 594 & 48470 & 138.48 \\
        \hline
        Others & Compositions not classified under predefined categories & 454 & 24912 & 69.13 \\
        \hline
        Unaccompanied Solo & Solo compositions for a single non-piano instrument & 207 & 3542 & 11.24 \\
        \hline
        Guitar Solo & Solo compositions for classical guitar & 192 & 1976 & 6.97 \\
        \hline
        Piano Trio & Compositions for piano, violin, and cello & 254 & 22736 & 68.51 \\
        \hline
        Organ Solo & Solo compositions for organ & 161 & 5923 & 20.01 \\
        \hline
        Piano Quintet & Compositions for piano and string quartet & 109 & 13382 & 34.69 \\
        \hline
        Piano Quartet & Compositions for piano, violin, viola, and cello & 84 & 9168 & 26.07 \\
        \hline
        Harpsichord Solo & Solo compositions for harpsichord & 84 & 17419 & 43.93 \\
        \hline
        Woodwind Ensemble & Ensembles consisting only of woodwind instruments & 63 & 3784 & 10.05 \\
        \hline
        Other Wind Ensemble & All kinds of wind ensembles beyond the woodwind family & 51 & 3206 & 8.06 \\
        \bottomrule
    \end{tabular}
\end{table*}

The category "Accompanied Solo" includes solo compositions for a single instrument with piano accompaniment. In contrast, the "Unaccompanied Solo" category refers to solo compositions for a single instrument without any piano accompaniment. The category "Other Wind Ensemble" includes all wind ensembles that are not exclusively composed of woodwind instruments, while "Woodwind Ensemble" consists exclusively of woodwind instruments. Compositions that do not belong to any of the predefined categories have been grouped under "Others".

\subsection{Metadata Standardization}

After extracting initial metadata, we performed additional standardization steps to ensure consistency across our dataset, particularly focusing on piece titles and composer names.

\subsubsection{Title Unification}
For composers with numerous pieces in our dataset, we identified potential duplicate entries by examining composer-instrumentation pairs. For the top 100 such pairs by frequency, we employed Claude-3.5-Sonnet to regenerate standardized titles using a carefully designed prompt. The model was instructed to group pieces with matching catalog numbers (BWV, K., Op., Hob, etc.) and movements, while maintaining separate entries for pieces with different catalog numbers or movements. For arrangements and transcriptions, we preserved both the original piece information and arranger attribution (e.g., original composer's piece with arrangement details), ensuring clear documentation of both the source material and its adaptation. All non-essential information such as performer names and years was removed to maintain consistency.

\subsubsection{Composer Name Normalization}
We observed multiple representation formats for composer names across video titles, including variations in name order (given name first vs. surname first) and completeness (full name vs. surname only). To address this, we implemented a systematic normalization process based on surname matching. By identifying all variants of each composer's name through case-insensitive surname searches, we created a comprehensive mapping of name variations to their standardized forms. This standardization was particularly crucial for prominent composers with numerous works in our dataset.

\subsection{Slide Segmentation Algorithm Details}\label{sec:slide-seg}
\subsubsection{Frame Analysis and Page List Generation}
The first step constructs a comprehensive page list through sequential frame analysis. The process begins by extracting frames at three-frame intervals and comparing each frame with its predecessor to detect visual changes. When differences are detected, the algorithm records the current frame index and classifies it as either \texttt{static} (stable page display) or \texttt{transition} (page turning animations such as crossfade and wipe effects) based on the previous frame's state. This frame-by-frame analysis produces a list of frame states and indices, which is then consolidated by merging consecutive frames of the same state. The resulting page list contains segments defined by their state (\texttt{static}/\texttt{transition}), start frame index, and end frame index. After generating this initial list, the algorithm removes all transition segments, retaining only the stable page displays for further processing.
\subsubsection{Audio-Image Pair Extraction}
The second step processes the filtered page list to extract corresponding audio-image pairs. For each static page segment identified in the previous step, the algorithm extracts both the stable frame image and its temporally aligned audio segment.
\subsubsection{Silent Segment Filtering}
The final step addresses segments containing silent audio, which commonly occur in specific parts of score videos. These silent segments are typically found in video introductions, outros, and transition screens between movements (distinct sections of a longer musical work). For example, a video of a multi-movement sonata might display a title page announcing "Movement II: Andante" before the music begins. By filtering out these silent segments, we ensure that our dataset contains only audio-image pairs with actual musical content.

\subsection{System and Staff Height Detection with Fine-tuned YOLOv8}\label{sec:lsyolo-system}

To crop and resize musical systems in sheet music images, we fine-tune two YOLOv8-medium\cite{yolov8} models for specialized detection tasks. Both YOLOv8-medium models are pretrained on COCO dataset\cite{cocodataset}. The first model, YOLO-system, detects and generates bounding boxes for musical systems within score pages. The second model, YOLO-staff, identifies staff regions containing key signatures and clefs in the leftmost measure of each cropped system.

For YOLO-system, we fine-tune the model on 1,306 labeled samples with a batch size of 24 for 5,400 iterations (100 epochs.) This model significantly improves upon our previous approach by effectively excluding non-musical elements while ensuring robust detection of complete systems, including notes that extend beyond the standard staff lines. Using the bounding boxes generated by the YOLO-system, we successfully crop individual systems from each score page.

YOLO-staff, fine-tuned on 1,254 labeled samples with a batch size of 24 for 5,200 iterations (100 epochs), enables precise detection of aforementioned specific staff regions within cropped systems from YOLO-system model's output. By measuring the staff height from these detections, we implement a standardization process that resizes all cropped systems to maintain a consistent staff height of 18 pixels. This approach provides a more reliable basis for size normalization compared to our previous rule-based method.

\subsection{Statistical Data Filtering Details}
\label{subsec:statistical_filtering}

\subsubsection{Pixel Intensity Metrics}

\begin{figure}[t]
  \centering
  \subfloat[]{\includegraphics[height=2.3cm]{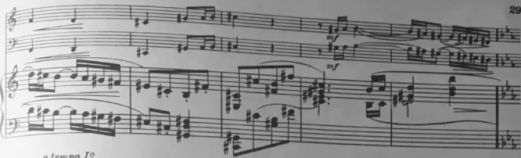}}\\[2pt]
  \subfloat[]{%
    \setlength{\fboxsep}{0pt}%
    \setlength{\fboxrule}{0.3pt}%
    \fbox{\includegraphics[height=2.3cm]{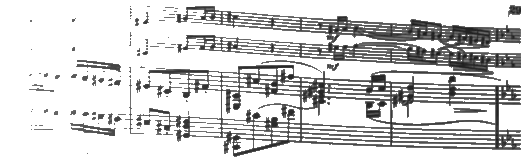}}%
  }
    \caption{Example of a poor-quality scan that necessitates our video-level intensity filtering. (a) Original score image showing low contrast with a grayish background. (b) The same image after our background normalization preprocessing (setting pixels above median-20 to white). Low-quality scans with uneven background colors often become illegible after normalization, demonstrating why we implement video-level intensity filtering to exclude scores with consistently poor scan quality from our dataset.}
    \label{fig:overundercrop}
\end{figure}

We implement comprehensive intensity-based filtering using both video-level and system-level metrics:

\paragraph{Video-Level Filtering}
For each video $v$, we compute the mean of median pixel values across all systems to identify videos with poor scan quality, which typically appear grayish due to low contrast:
\begin{equation}
   \textit{video\_intensity}_v = \frac{1}{N}\sum_{i=1}^{N} \textit{median}(\textit{pixels}_i)
\end{equation}
where $N$ is the number of systems in video $v$. Videos with $\textit{video\_intensity}_v < 200$ are excluded.

\paragraph{System-Level Anomaly Detection}
To detect systems with inverted colors (commonly used to indicate repeated sections), we compute a pixel anomaly score for each system $i$ in a video:
\begin{equation}
   \textit{median\_anomaly}_i = \frac{\textit{median}(\textit{pixels}_i) - \textit{median}(\textit{pixels}_\textit{video})}{\textit{median}(\textit{pixels}_\textit{video})}
\end{equation}
\begin{equation}
   \textit{mean\_anomaly}_i = \frac{\textit{mean}(\textit{pixels}_i) - \textit{mean}(\textit{pixels}_\textit{video})}{\textit{mean}(\textit{pixels}_\textit{video})}
\end{equation}
\begin{equation}
   \textit{pixel\_anomaly\_score}_i = \frac{\textit{median\_anomaly}_i + \textit{mean\_anomaly}_i}{2}
\end{equation}
Systems with $|\textit{pixel\_anomaly\_score}_i| > 0.1$ are filtered out.

\subsubsection{Dimensional Analysis}
We implement three types of dimensional constraints:

\paragraph{Basic Height Constraints}
Systems must satisfy:
\begin{equation}
   70 \leq \textit{height} \leq 390 \text{ pixels}
\end{equation}
\begin{equation}
   \textit{height} < \textit{width}
\end{equation}

\begin{table*}[ht]
    \centering
    \caption{List of the videos in YTSV-T11}

    \begin{tabular}{l l l c}
        \toprule
        \textbf{Composer} & \textbf{Work} & \textbf{YouTube ID} & \textbf{Included Segments} \\
        \midrule
        Clara Schumann & Piano Sonata in G minor & \texttt{Pw4fMNMO90U} & 74 \\
        Friedrich Gulda & Prelude and Fugue & \texttt{V2h23Dsw57A} & 31 \\
        Alexander Borodin & Petite Suite & \texttt{7vBkBCa3n4o} & 31 \\
        Lev Abeliovich & 5 Pieces for Piano & \texttt{07zYLY1YTj0} & 29 \\
        Carl Czerny & Studio No. 29 Op. 409 & \texttt{J5WRTAYtaOg} & 24 \\
        Mily Balakirev & Mazurka No. 5 & \texttt{BqAWfT76pJY} & 19 \\
        Alexander Borodin & Petite Suite & \texttt{38P9U3WRX9w} & 15 \\
        Giovanni Sgambati & Vecchio minuetto & \texttt{EXNifef40vU} & 14 \\
        Giovanni Sgambati & 2 Concert Etudes & \texttt{olBJh\_5rv2c} & 10 \\
        Carl Czerny & Album élégant des Dames Pianistes Vol.3 & \texttt{n\_Sn48u1t94} & 3 \\
        Carl Czerny & Romance Op. 755 No. 12 & \texttt{4Pa4x9SDNdw} & 1 \\
        \bottomrule
    \end{tabular}
    \label{tab:composer_list}
\end{table*}

\paragraph{Height Anomaly Detection}
To identify cases of over-detection (multiple systems detected as one) or partial detection (incomplete system capture), we compute for each system $i$ in a segment $s$:
\begin{equation}
   \textit{height\_anomaly\_score}_i = \frac{\textit{height}_i - \mu_{\textit{height}_s}}{\sigma_{\textit{height}_s}}
\end{equation}
where $\mu_{\textit{height}_s}$ and $\sigma_{\textit{height}_s}$ are the mean and standard deviation of system heights in segment $s$. Systems with $|\textit{height\_anomaly\_score}_i| > 1.2$ are excluded.

\paragraph{Overlap Analysis}
Using the bounding box coordinates predicted by YOLOv8-System, we compute the overlap score for each pair of systems $(i,j)$:
\begin{equation}
   \textit{overlap\_score}_{i,j} = \frac{\textit{intersection\_area}_{i,j}}{\min(\textit{height}_i, \textit{height}_j)} \times \max(\textit{height}_i, \textit{height}_j)
\end{equation}
Systems with $\textit{overlap\_score} > 25$ are filtered out.

\subsubsection{Temporal Constraints}
Each segment duration $d$ must satisfy:
\begin{equation}
  3 \leq d \leq 20 \text{ seconds}
\end{equation}
This filtering serves as an additional safeguard to remove remaining noise segments that persist after silent segment filtering. While most video introductions, outros, and chapter titles are caught by silence detection, this duration-based constraint helps eliminate any remaining problematic segments while also excluding segments that exceed our model's processing capacity.

\subsection{Test Set: YTSV-T11}
The test set(YTSV-T11) used in the image-to-audio generation is provided in \cref{tab:composer_list}.

\section{Resident Vector Quantization Details}
\subsection{RQVAE Model Adaptations for Sheet Music Processing}
\subsubsection{Core Architecture Modifications}
The original RQVAE architecture was designed for RGB images with three channels. Our adaptation processes single-channel grayscale sheet music images, significantly reducing model complexity. We employ four unshared codebooks, each containing 1024 codes, with a model dimension of 256.

\begin{figure*}[ht]
    \begin{center}
    \includegraphics[width=1\textwidth]{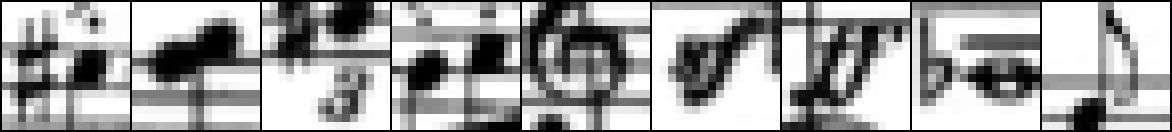}
    \caption{Example patches showing the 16×16 pixel resolution of individual tokens}
    \label{fig:rqvae_patch}
    \end{center}
\end{figure*}

\subsubsection{Compression Strategy}
While the original model achieved 32x compression using a channel multiplication sequence of \texttt{[1, 1, 2, 2, 4, 4]}, we implement a modified sequence of \texttt{[1, 1, 2, 2, 4]} for 16× compression. This ensures each token captures features at a sub-staff-line scale, crucial for precise musical notation representation.

\subsubsection{Architectural Refinements}
We removed attention blocks from both the encoder's final downsampling layer and the decoder's initial upsampling layer. The attention blocks in the original model help maintain global coherence across larger image regions. However, in our case, we aim to tokenize sheet music at a very local level, focusing on fine details like individual note heads and staff lines. Since our tokens need to capture these small-scale features rather than long-range dependencies, we found that attention blocks were not only unnecessary but potentially detrimental to our specific use case. Instead of using the perceptual loss implemented in VQ-GAN\cite{esser2020taming}, we used MSE loss between the ground truth and reconstruction after normalizing and weighting the activation values between convolutional layers of the Zeus OMR\cite{mayerZeus} model encoder that we reproduced, following the approach proposed in \cite{rodriguez2023ocr}.

\subsubsection{Resolution-Adaptive Training}

\begin{figure*}[ht]
    \begin{center}
    \includegraphics[width=1\textwidth]{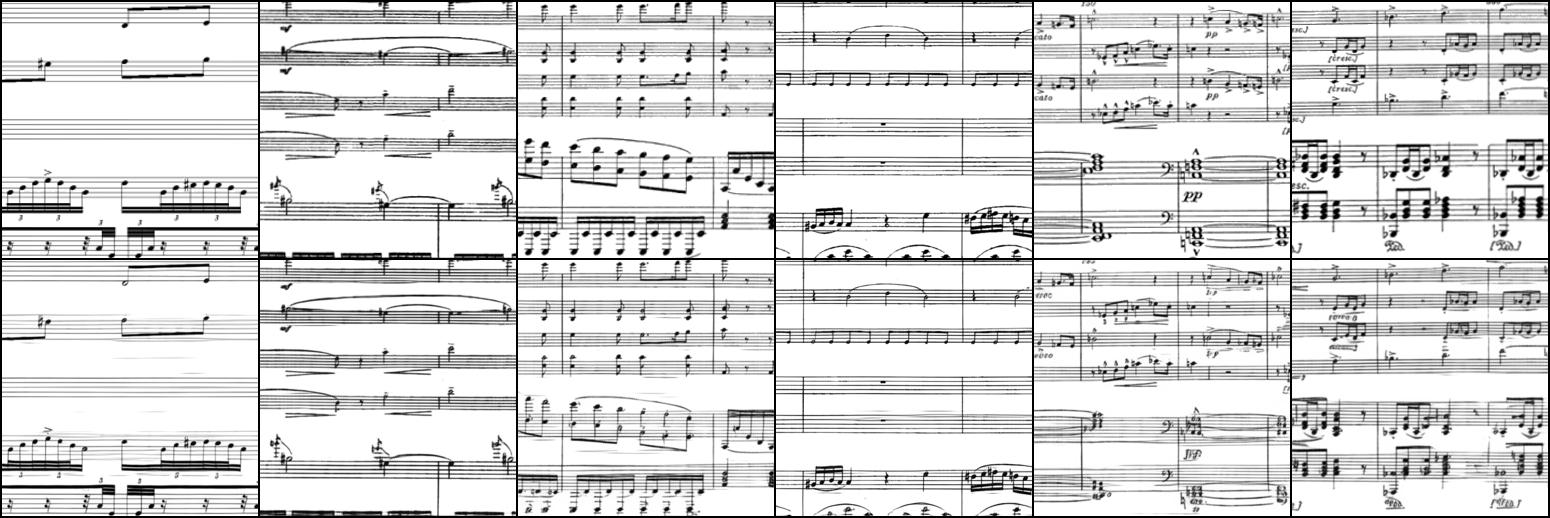}
    \caption{Comparison between input sheet music (top) and model reconstruction (bottom), demonstrating reconstruction artifacts in staff lines when a model trained only on 64×64 pixel image crops processes 256×256 pixel inputs(unseen image size).}
    \label{figure-recons}
    \end{center}
\end{figure*}

Sheet music presents unique challenges in image dimensions, varying from compact piano scores (around 70 pixels in height) to extensive multi-staff compositions. To effectively tokenize these diverse formats, we needed to address a key limitation in the original RQVAE model, which was trained on fixed resolution images and failed to maintain consistent spatial relationships when processing larger inputs. Our solution implements a dynamic training strategy with resolution-specific parameters:

\begin{itemize}
    \item 70-130 pixels: 64-pixel random crops, batch size 256
    \item 130-260 pixels: 128-pixel random crops, batch size 128
    \item 260-360 pixels: 256-pixel random crops, batch size 32
    \item 360-390 pixels: 352-pixel random crops, batch size 16
\end{itemize}

\vfill

\end{document}